\documentclass[twocolumn,twocolappendix,trackchanges]{aastex631}

\shorttitle{}
\shortauthors{}
\graphicspath{{./}{fig/}}

\newcommand{\luB}{Luhman 16 B}

\usepackage{amsmath}
\usepackage{graphicx}
\usepackage{siunitx}
\usepackage{multirow}
\usepackage{booktabs}
\usepackage{vcell}
\usepackage{subfigure}
\usepackage{tikz}
\usepackage{natbib}

\begin{document}

\title{Latitude-dependent Atmospheric Waves and Long-period Modulations in Luhman 16 B from the Longest Lightcurve of an Extrasolar World}
\shorttitle{Short- and Long-period Lightcurve Evolution in Luhman 16 B}

\author[0000-0002-6372-8395]{Nguyen Fuda}
\affiliation{Lunar and Planetary Laboratory, University of Arizona, 1640 E. University Boulevard, Tucson, AZ 85721, USA}
\email{fuda@lpl.arizona.edu}

\author[0000-0003-3714-5855]{Dániel Apai}e
\affil{Steward Observatory, The University of Arizona, 933 N. Cherry Avenue, Tucson, AZ 85721, USA}
\affiliation{Lunar and Planetary Laboratory, University of Arizona, 1640 E. University Boulevard, Tucson, AZ 85721, USA}

\author[0000-0003-1149-3659]{Domenico Nardiello}
\affil{INAF-Osservatorio Astronomico di Padova, Vicolo dell'Osservatorio 5, I-35122 Padova, Italy}

\author[0000-0003-2278-6932]{Xianyu Tan}
\affil{Tsung-Dao Lee Institute, Shanghai Jiao Tong University, 520 Shengrong Road, Shanghai, People’s Republic of China}
\affil{School of Physics and Astronomy, Shanghai Jiao Tong University, 800 Dongchuan Road, Shanghai, People’s Republic of China}

\author[0000-0001-7356-6652]{Theodora Karalidi}
\affil{Department of Physics, University of Central Florida, 4111 Libra Dr, Orlando, FL 32816, USA}

\author[0000-0003-4080-6466]{Luigi Rolly Bedin}
\affil{INAF-Osservatorio Astronomico di Padova, Vicolo dell'Osservatorio 5, I-35122 Padova, Italy}

%
%

\accepted{February 21, 2024}
\revised{January 26, 2024}
\received{November 8, 2023}


\begin{abstract}

In this work, we present the longest photometric monitoring of up to 1200 hours of the strongly variable brown-dwarf binaries Luhman 16 AB and provide evidence of $\pm5\%$ variability on a timescale of several-to-hundreds of hours for this object. We show that short-period rotational modulation around 5 hours ($k=1$ wavenumber) and 2.5 hours ($k=2$ wavenumber) dominate the variability under 10 hours, where the planetary-scale waves model composed of $k=1$ and $k=2$ waves provides good fits to both the periodogram and light curve. In particular, models consisting of three to four sine waves could explain the variability of light curve durations up to 100 hours. We show that the relative range of $k=2$ periods is narrower compared to $k=1$ period. Using simple models of zonal banding in Solar System giants, we suggest that the difference in period range arises from the difference in windspeed distribution at low and mid-to-high latitudes in the atmosphere. Lastly, we show that Luhman 16 AB also exhibits long-period $\pm5\%$ variability with periods ranging from 15 hours up to 100 hours over the longest monitoring of this object. Our results on $k=1$ and $k=2$ waves and long-period evolution are consistent with previous 3D atmosphere simulations, demonstrating that both latitude-dependent waves and slow-varying atmospheric features are potentially present in Luhman 16 AB atmospheres and are significant contribution to the light curve modulation over hundreds of rotations.
\end{abstract}

\keywords{Brown dwarfs (185) -- Atmospheric circulation (112) -- Broad band photometry (184) -- Exoplanet atmospheres (487)}

\section{Introduction} \label{sec:Intro}





Most, if not all, brown dwarfs harbor heterogeneous condensate clouds \citep[][]{buenzli_brown_2014,metchev_weather_2015}. Although spatially unresolved, cloud cover has been successfully characterized through time-resolved observations of the rotating atmospheres \citep[e.g.,][]{biller_time_2017,artigau_variability_2018}. Early spectrophotometric observations showed that the rotational modulations are caused by cloud thickness variations \citep[][]{radigan_large-amplitude_2012,apai_hst_2013}, and that these cloud decks  have sustained three-dimensional (longitudinal-vertical) structures \citep[][]{buenzli_VerticalAtmosphericStructure_2012,yang_extrasolar_2016}. 

Prior to 2017, no long-term infrared monitoring data was available. Short-term (few hours-long) photometric light curves could be fitted well via elliptical spots \citep[e.g.,][]{apai_hst_2013,karalidi_aeolus_2015}. However, the long-term Spitzer monitoring -- which became available in 2017 through a dedicated large Spitzer program -- revealed complex and continuously evolving light curves \citep[][]{apai_zones_2017}. The nature of the evolving light curves was inconsistent with and could not be fitted by elliptical spots: Rather, \citet[][]{apai_zones_2017} showed that the light curves are very well reproduced by a new model: planetary-scale waves. In this model, planetary-scale waves trapped in zonal circulation modulate cloud thickness which, in turn, produces rotational modulation in the rotating atmospheres. The initial study, however, was limited to continuous observations of only four rotations, and only on three L/T dwarfs (although each object was visited eight times). 


In 2021, \citet[][]{apai_tess_2021} presented a dataset that covered 20$\times$ more rotational periods continuously than any previous dataset. These TESS \citep[][]{ricker_transiting_2015} light curves probed rotational modulations in Luhman 16 AB -- an L7.5+T0.5 binary brown dwarfs system that is also the closest to Earth. The Luhman 16AB light curve showed complex evolution over 100 rotations. These changes were successfully modeled with the planetary-scale wave model \citep[][]{apai_tess_2021}, just like previously the Spitzer light curves. However, the longer light curves allowed more comprehensive analysis and revealed the presence of not one, but multiple similar rotational periods. The authors attributed the range of periods observed to zonal circulation and differential rotation, i.e., wave-modulated cloud structures trapped in zones at different latitudes will have slightly different periods due to differential rotation.

Studies using other techniques also found evidence for zonal circulation and planetary-scale waves: \citet{millar-blanchaer_detection_2020} used VLT time-resolved polarization measurements to constrain the surface brightness distribution in the atmosphere of Luhman 16A and B. They found evidence for asymmetry (net polarization signal) which is consistent with the presence of bands and zones. \citet{mukherjee_modeling_2021} also reproduced the measured polarization of Luhman 16AB using polarization radiative transfer modeling that is coupled with general circulation model (GCM) outputs, and those GCM exhibits enhanced cloud abundances near the equatorial zones. Targeting another brown dwarf with high-amplitude rotational modulations, \citet[][]{zhou_roaring_2022} also found that its light curve is evolving rapidly but not irregularly, and showed that the modulations can be well fit with the planetary-scale waves model. The presence of zonal circulation and planetary-scale waves should come as no surprise, as they are present in most, if not all, planetary atmospheres in the Solar System, including Earth and Jupiter \citep[e.g.,][]{fletcher_how_2020}. Zonal circulation and jets are predicted in rotation-dominated brown dwarf atmospheres \citep[][]{zhang_atmospheric_2014,showman_atmospheric_2019, tan_jet_2022, hammond_shallow-water_2023} and planetary-scale waves are also seen in the most comprehensive general circulation models \citep[e.g.,][]{tan_atmospheric_2021}
We note that one prominent measurement \textit{seemingly} contradicts the presence of bands and zones in Luhman 16AB: \citet[][]{crossfield_global_2014}, which inverted time-resolved CO line profile modulations via a method developed for starspots, to identify the most likely surface brightness distribution at the CO-probed low-pressure region. The surface brightness model from this study does not show evidence for bands and zones -- but due to its nature, the method is insensitive to such structures (see supplementary online material of \citealt[][]{crossfield_doppler_2014}). \citet[][]{karalidi_maps_2016} showed that the same data, when considering the uncertainties does, in fact, not contradict surface brightness models derived from other methods. 

Arguably, important advances over the past ten years have been due to the improving temporal coverage and precision of time-resolved observations of rotational modulations. Partial- or single-rotation modulations do not provide enough constraints, due to the information loss inherent to hemisphere-integrated signal and the inverse problem of mapping exoplanets \citep[e.g.,][]{cowan_light_2013}. Modulations over four back-to-back rotations (in 32 epochs) provided strong evidence \textit{against} elliptical spots playing dominant roles in shaping light curves \citep[][]{apai_zones_2017}. A hundred-rotation coverage provided by TESS \citet[][]{apai_tess_2021} allowed initial characterization of the zones, wind speeds, and differential rotation in a brown dwarf atmosphere \citep[][]{apai_tess_2021}, and the tentative identification of $k=1$ as well as $k=2$ waves.

However, changes over very long timescales (thousands of rotations) and periods shorter than the rotational period could not be studied due to the lack of data. Our paper presents a new, longest monitoring dataset of brown dwarf atmospheres thus far that provides higher cadence data and allows detailed characterization of the $k=2$ waves for the first time. Furthermore, we also offer very long baseline data, which enables the exploration of changes on timescales well exceed the rotational modulations -- and therefore give access to a yet unexplored region of new atmospheric processes in brown dwarf atmospheres.

This paper extends the methodology of \citet[][]{apai_tess_2021} and uses higher-cadence and longer-baseline data to provide improved analysis of short- and long-period modulations. The paper is organized as follows: In Section \ref{sec:TESSData}, we discuss the TESS monitoring of Luhman 16 AB in Sectors 36-37 and strategies implemented to assess photometric contamination. In Section \ref{sec:Periodo} we discuss our generalized Lomb-Scargle periodogram to explore the temporal structure and period components in the variability.  In Section \ref{sec:LCfit}, we describe a multi-sine, planetary-scale waves model to explain the behavior of the light curve in short periods for periods under 10 hours. In Section \ref{sec:LongPeriodLCevolution}, we will discuss the evolution of the long-period light curve from 15-100 hours. In Section \ref{sec:Discussion}, we discuss the interpretation of short-period components, provide a tentative comparison with the Solar System gas giants toy model, and discuss the long-period results in the context of 3D atmosphere studies. Finally, we summarize our key findings in Section \ref{sec:Conclusion}.

\begin{figure*}[ht!]
\label{fig_photom_lc}
\centering
\includegraphics[width=0.95\textwidth]{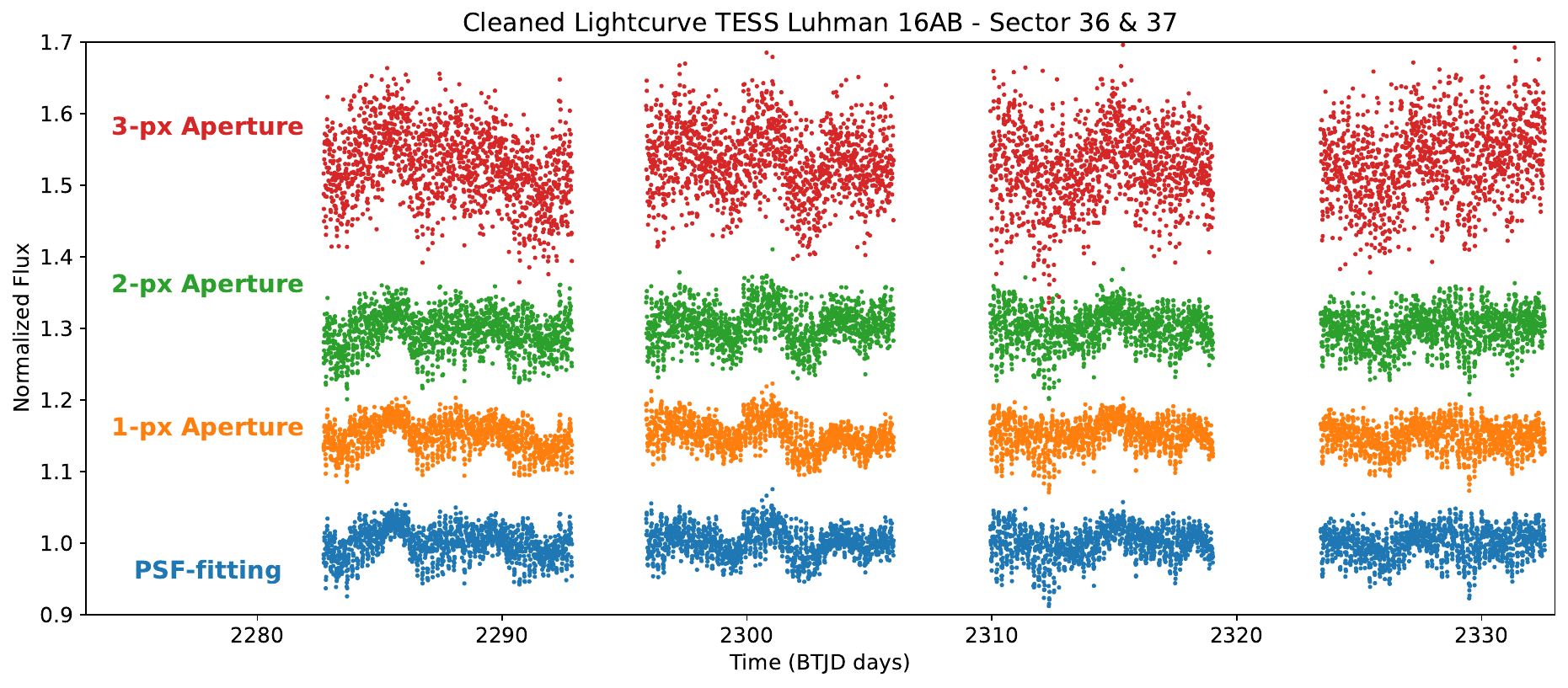}
\caption{Significant variability exists in TESS 50-day light curve of Luhman 16AB in Sector 36 and 37. The cadence is 10 minutes. Gaps in data are due to downlink periods. Different curves show the result of aperture size adjustment from the PATHOS light curve extraction pipeline. The PSF-extracted light curve has the smallest photometric error.}
\end{figure*}

\section{TESS Photometry} \label{sec:TESSData}

In this paper, we are presenting TESS \citep[][]{ricker_transiting_2015} photometry data (600--1,000 nm bandpass) of Luhman 16AB in Sectors 36 and 37. Photometry data is collected from March 07 2021 to April 28 2021 in two consecutive TESS orbits, covering a baseline of more than 50 days -- about 48 of which are continuous science data not interrupted by data downlink gap. The downlink period totaled 2.12 days during two TESS orbits for sectors 36 and 37\footnote{\href {https://heasarc.gsfc.nasa.gov/docs/tess/cycle3_drn.html}{TESS Cycle 3 Data Release Notes}}. 

We used the PATHOS pipeline from \citet{nardiello_psf-based_2020} for the extraction and correction of the light curves from TESS full-frame images. PATHOS is a pipeline designed to extract high-precision photometric products for objects in crowded star fields using empirical PSFs and subtraction of neighbors. This helps to minimize potential flux contamination for faint objects. 

The photometry presented in this work has a 10-minute cadence compared to a cadence of 30 minutes in Sector 10 data from \citet{apai_tess_2021}. Figure \ref{fig_photom_lc} shows the full light curve of Luhman 16AB extracted with PATHOS with different photometric extraction apertures: the point-spread-function (PSF), and circular aperture with radii of 1, 2, and 3 pixels. Photometric noise becomes more significant as the aperture radius increases. In order to minimize photometric noise, we primarily use the PSF-aperture-extracted light curve throughout this study.

The average photometric error is 4.5\%. The Earth and the Moon enter the field of view at the start of every orbit and introduce a significant amount of scattering photons. Background sources' scattered light is filtered out by removing photometric points with local sky noise factor $\sigma_{\text{SKY}}>140$ $e\:s^{-1}$ and bad quality flag \verb|DQUALITY|$\neq0$. This is in order to remove noisy photons from background sources, particularly from scattered lights of the Earth and Moon. 

The temporal data itself is contaminated by various instrumental and astrophysical artifacts. We assessed possible sources of instrumental and astrophysical contamination: A) the spacecraft positional fluctuation -- how the spacecraft pointing fluctuates on a pixel-by-pixel scale during observations; and B) background sources variations. The window function contamination, e.g., how the gaps in data sampling bias acquisition of certain periods of variability, will be discussed more in detail in Section \ref{sec:Periodo}.




\begin{figure*}[htp!]
\centering
\includegraphics[width=0.98\textwidth]{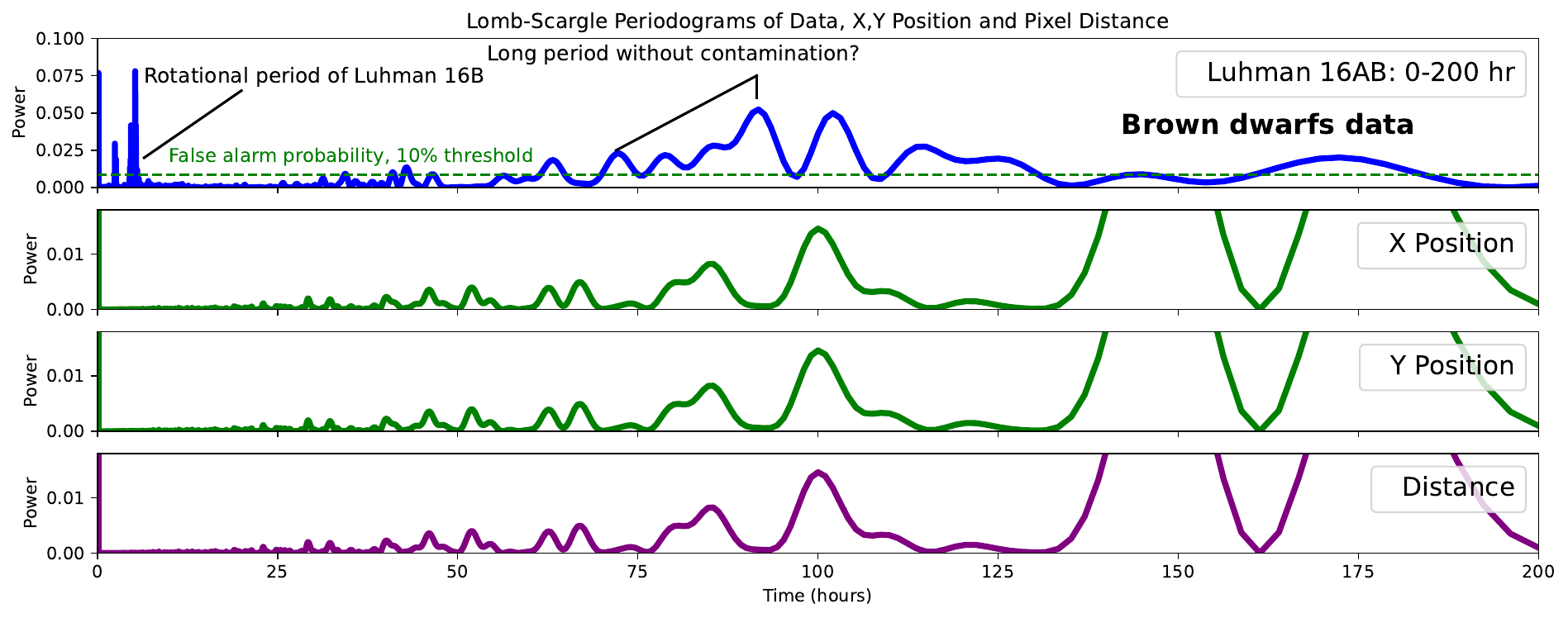}
\caption{The GLS (Generalized Lomb-Scargle) periodograms of the TESS data (blue, first panel) and the instrumental pointing accuracy: X \& Y position through time (green, second and third panel), and drift distance from the median center pixel through time (purple, fourth panel). The data periodogram shows strong power under 10 hours, which does not appear in the instrumental pointing over time. {Peaks in the data curve which correspond to troughs in the spacecraft positional variation curves suggest that this period range is likely not contaminated, and vice versa. The green dashed line indicated the false-alarm-probability (FAP) level with a $10\%$ power threshold, where peaks under the FAP level are potentially spurious.}
\label{fig:XYposDistance_periodogram}}
\end{figure*}

The centroid of the PSF extraction on Luhman 16AB fluctuates on a pixel-by-pixel scale due to spacecraft jitters that introduce sources of periodic contaminants into data. The gradual, slight motion of the photometric aperture across the starfield might introduce a trend between pointing position and measure intensity. 

Thus, we analyzed the time variation of these centroid positions and their distance to the actual source with periodograms, as shown in Figure \ref{fig:XYposDistance_periodogram}.  In order to quantify the relative strength of periodic components in the time series data,  we use the Lomb-Scargle (GLS) periodogram method to transform time-series data into the frequency domain and explore their period distribution. 

Slight spacecraft pointing drifts lead to variations in the celestial (X/Y) pointing positions and distance to the source. These variations are small compared to short-period brightness changes as shown in Figure \ref{fig:XYposDistance_periodogram}. This is true for powers with periods below 25 hours but is not valid for powers with longer periods. The degree to which long-period powers in the data are affected is not uniform: gaps in the positional variation powers sometimes coincide with peaks in the data (i.e., around 72 and 90 hours). For periods larger than 125 hours, particularly from 150-200 hours, the periodogram powers of the  XY positional variations are at least twice as large compared to powers at 100 hours. To avoid potential contamination, we take a 100-hour period to be the upper limit when assessing the long-period variation in the light curve in Section \ref{sec:LongPeriodLCevolution}.

Another potential source of long-period contamination is the temporal variation of background sources. TESS has a pixel edge of 21 arcseconds -- thus neighboring sources that fall into a region of $\pm1$ pixel or 20--40 arcseconds (1--2 pixels) will potentially contaminate the signal. Past work of \citet{apai_tess_2021} has shown that no bright stars are in the vicinity of Luhman 16 And that the strongest signal with the highest amplitude comes from the aperture centered around Luhman 16. They used deep HST multi-epoch photometry to assess the brightnesses, variability amplitudes, and distances of background stars around Luhman 16AB. By combining 12 HST epochs data, they showed that none introduced amplitude similar to that of Luhman 16AB to the TESS data.


\section{Periodogram Analysis} \label{sec:Periodo}

The generalized Lomb-Scargle (GLS) periodogram is a method to efficiently calculate the Fourier power spectrum estimator using unevenly sampled data, providing a way to understand the underlying periods of wave-like signals. Here, we used the GLS periodograms to explore the contribution of each periodic component to the overall power spectrum of the data. In the following sections, we compare the power spectrum of all time-varying components, including the reduced photometric data, the window function (which describes data collection windows), and Luhman 16AB coordinates given in the TESS detector pixel coordinates. We then generated synthetic fits for the periodogram of the data to understand the makeup of its variable components.

\begin{figure*}[ht!]
\centering
\includegraphics[width=\textwidth]{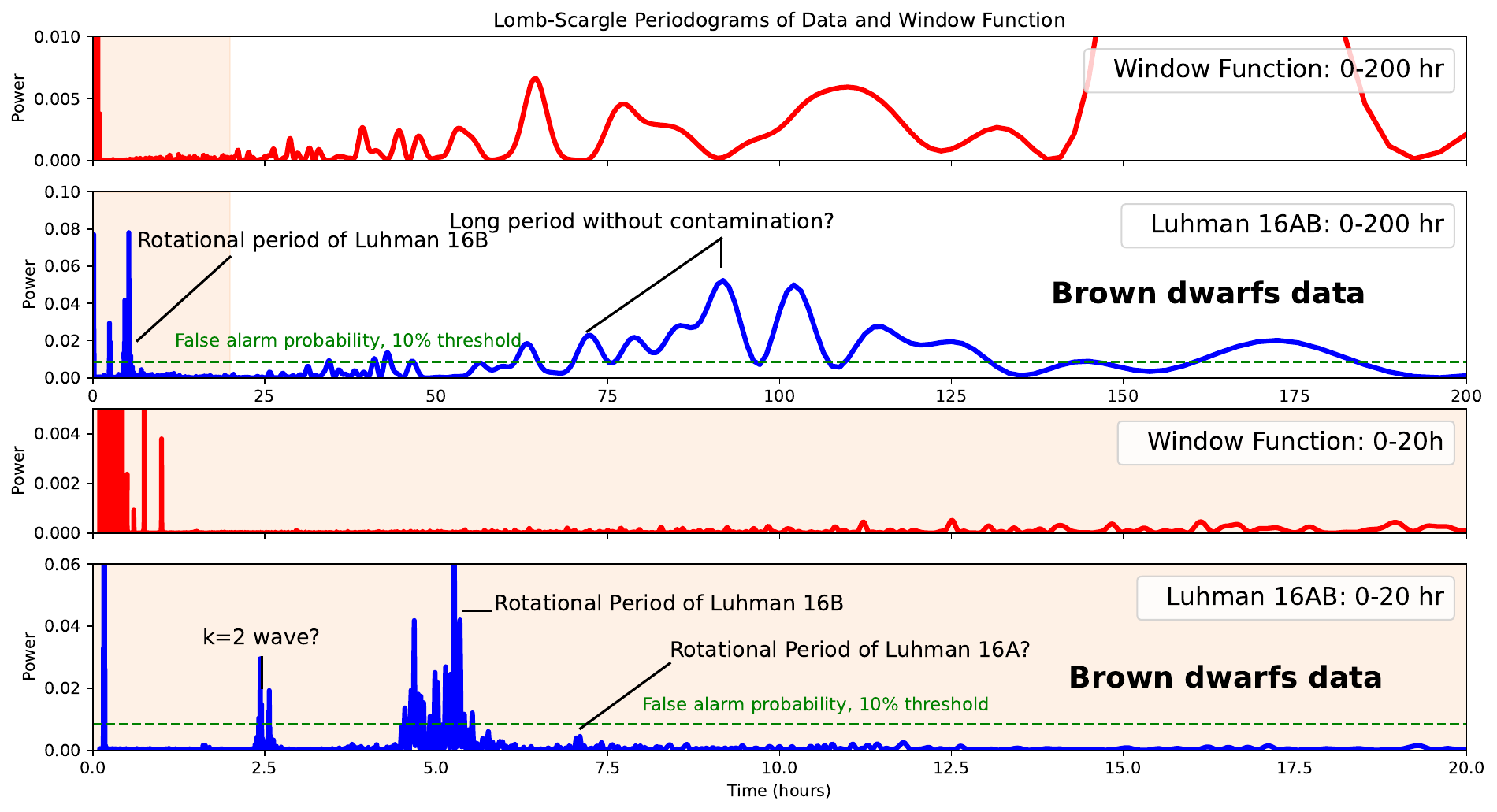}
\caption{The GLS periodograms of the window function describing gaps in data collection (red) and the data (blue). Two upper panels are for the period range from 0--200 hours. Two lower panels are for the period range from 0--20 hours. Little power contamination to the data exists for periods shorter than 20 hours while contamination is significant for longer periods. Inset texts indicate the literature value for the rotational periods of Luhman 16 A ($\approx7$-hour) and B ($\approx5$-hour), as well as the potential $k=2$ wavenumber periods around 2.5 hours. {Peaks in the data curve which correspond to troughs in the window function curves suggest that the corresponding period range is not contaminated, and vice versa. The green dashed line indicated the false-alarm-probability level with a $10\%$ power threshold.}}
\label{fig:periodogram_windowFunction_Data}
\end{figure*}

\begin{figure*}
    \centering
\includegraphics[width=\textwidth]{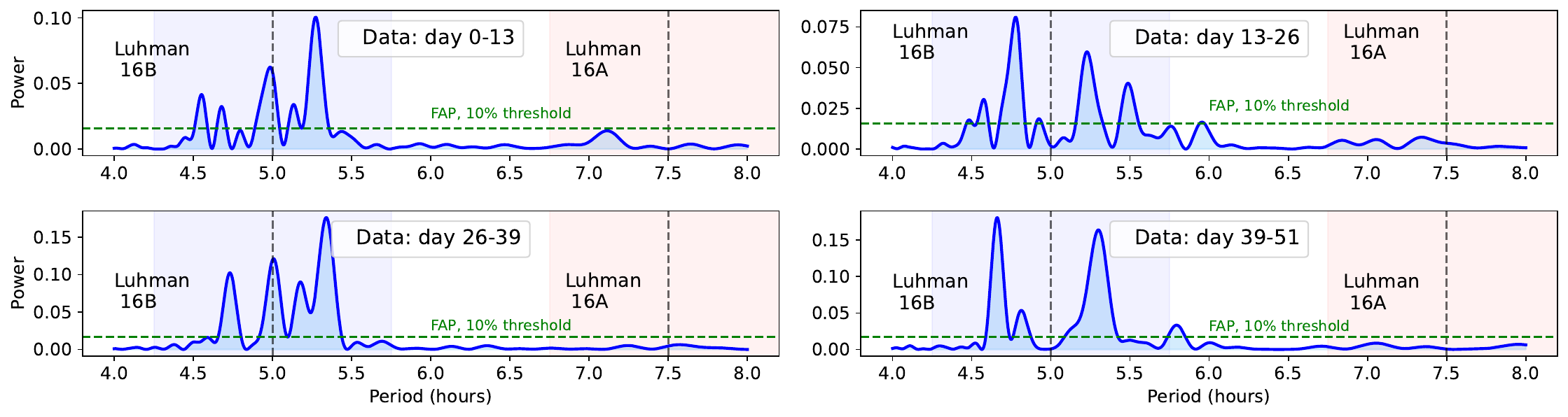}
\caption{{The periodograms of every consecutive 13-day data collection segment that make up the overall 52-day dataset. The dashed line indicated the rotational periods of Luhman 16 A ($\sim$ 7.5 hours) and Luhman 16 B ($\sim$ 5 hours). The colored boxes indicate the $\pm 1$ hour period range. Compared to Luhman 16 B, the power around the rotational period of Luhman 16 A is very low and stays flat throughout the data collection. A false-alarm-probability (FAP) level with a $10\%$ power threshold is shown.}}
\label{fig:periodogramCompare_A_and_B}
\end{figure*}

\begin{figure*}[htp!]
\centering
\includegraphics[width=\textwidth]{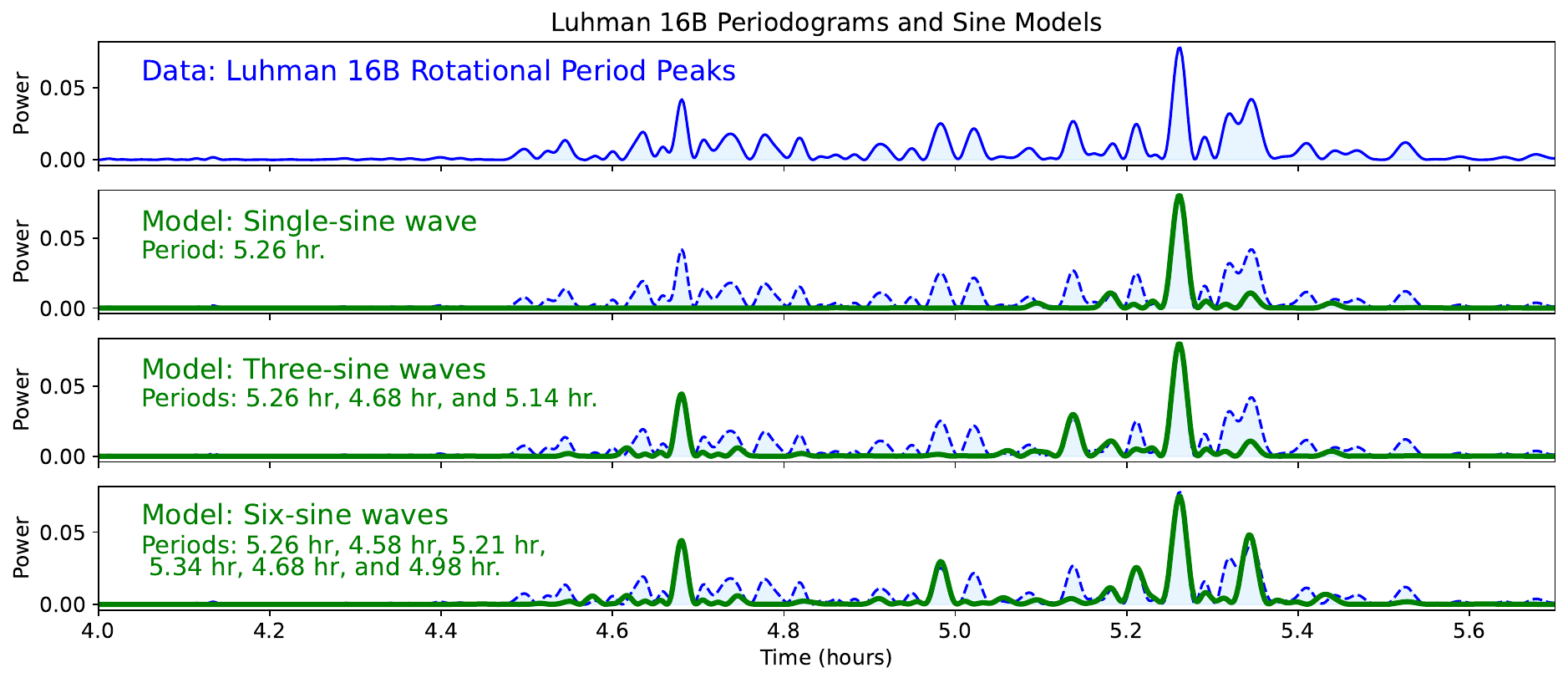}
\caption{{Periodogram fits for Luhman 16 B rotational period around 5 hours using different multi-sinusoidal models. For frequency-domain analysis, phase information is not needed, only the periods and amplitude. For each model plot, the best-fit periods are notated in the legend. One-sine, three-sine, and six-sine models (green curves) were used to fit the periodogram of the data (blue curve) following \citet{apai_tess_2021}. This periodogram fits show that a three-sine model is the simplest model that captures the highest amplitude structures (at periods of 4.7-hour and 5.25-hour) around the Luhman 16 B rotational period of 5 hours. }\label{fig:periodogram_fits}}
\end{figure*}

\subsection{Comparing time-varying components with the GLS periodograms }\label{subsec:periodogramMethod}



Due to the nature of our long baseline data extending to about 50 days ($\sim$ 1200 hours), variability may be seen from short (under 20 hours) to longer timescales (up to 200 hours). We first created GLS periodograms to compare the long-period variability of the light curve with the positional variation of TESS, which is shown in Figure \ref{fig:XYposDistance_periodogram}. {To distinguish between real and spurious peaks, we added the false alarm probability (FAP) with a power threshold of 10\%. The same threshold of FAP is applied through the data periodograms in Figure \ref{fig:XYposDistance_periodogram}, \ref{fig:periodogram_windowFunction_Data} and \ref{fig:periodogramCompare_A_and_B}. It can be seen that peaks around the rotational period of Luhman 16 B (5-hour), as well as long-period peaks from 70-125 hours, are well above the FAP level.} 

The positional variations are an important metric to assess pointing accuracy and to rule out time-dependent photometric contamination coming from spacecraft jitters or position changes. Positional variations are analyzed using the X, and Y pixel coordinates of Luhman 16AB derived from the TESS full-frame image. The distance drift is calculated using the median coordinates of all frames. 


The window function is an important metric to assess the reliability of the periodogram signal against potential sampling biases. To assess potential contamination from sampling, we created GLS periodograms to compare at the same time the periodicity of the light curve and the window function. In Figure \ref{fig:periodogram_windowFunction_Data}, the GLS periodogram of the data and the window function are shown. Upper panels are for long periods of up to 200 hours and lower panels are for short periods of up to 20 hours.

To create the window function, we generated an evenly-spaced array with a resolution equal to the 10-minute cadence of the TESS data. The array is defined to be 1 where data exists and 0 elsewhere. Generally, the window function shows four significant gaps of no data collection, corresponding to a total of four data down-link gaps in the two TESS orbits. The ability to capture robust periodogram power will be inhibited for periods similar to or larger than the timescale of these gaps in the window function.

\subsection{Synthetic sine fits for the periodogram}\label{subsec:periodoSyntheticSineFit}


\citet{apai_tess_2021} showed that the periodogram calculated from shorter TESS data segments ($\sim$500 hours) can be well fit by the planetary-scale wave model \citep[][]{apai_zones_2017}. We also test this model on our longer and higher-cadence data, following the synthetic periodogram fit outlined in \citet{apai_tess_2021}. Our goal was to find a linear combination of sine waves whose periodogram matches the observed pattern. We generated synthetic light curves composed of one-, three- and six-period multi-sine models in the form $\Sigma_i \alpha_i \sin (2\pi \omega_i t)$ and added random, uniformly distributed noise matching the amplitude of the noise in the TESS data. For each of these models, we searched the parameter space for a best-fit solution with Sequential Least-Squares Programming gradient fitting\footnote{\href{https://docs.scipy.org/doc/scipy/reference/generated/scipy.optimize.minimize.html}{\textbf{scipy.optimize.minimize} Documentation}}. For frequency-domain analysis, phase information is not needed and therefore not included. The fit results are shown in the second, third, and fourth panels of Figure \ref{fig:periodogram_fits}.

\vspace{-2mm}
\begin{figure*}[htp!]
\subfigure[Luhman 16 GLS Periodograms from TESS Light Curve]{
    \centering
    \includegraphics[width=0.6\textwidth]{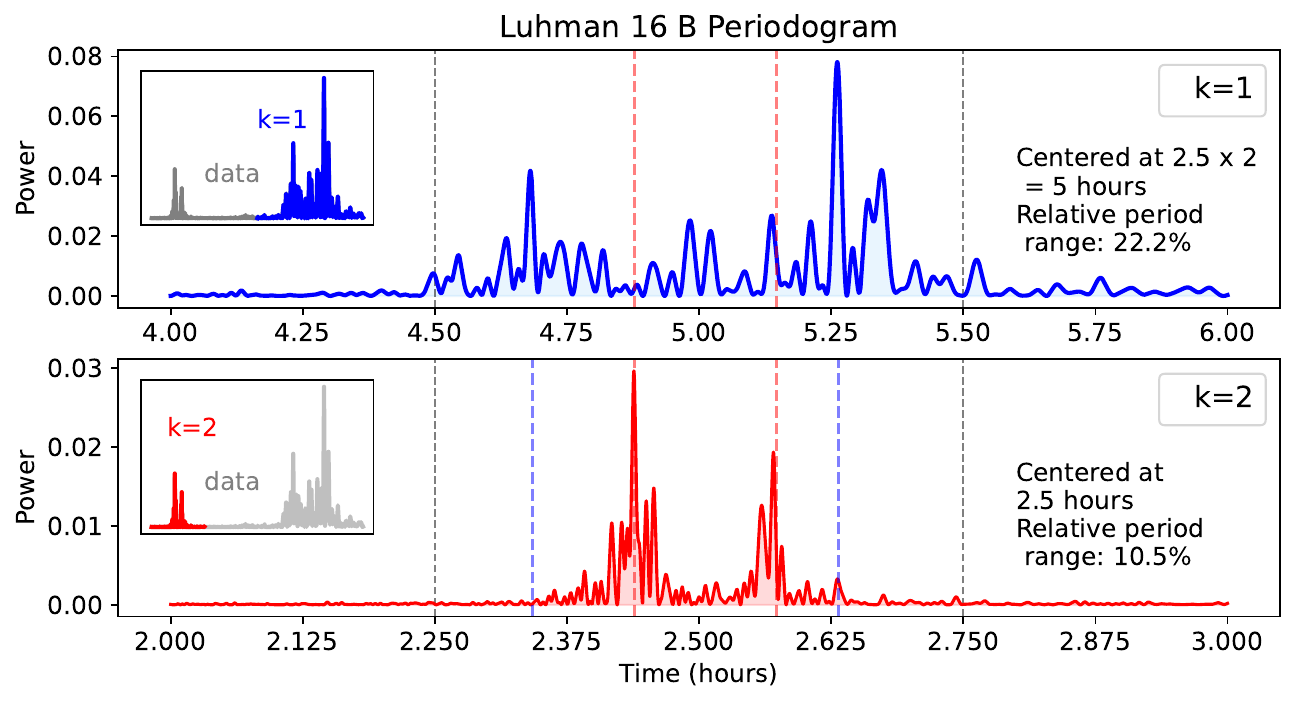}}
\quad
\subfigure[Neptune Power Spectra from \textit{Kepler/K2} light curve]{
    \centering
    \includegraphics[width=0.335\textwidth]{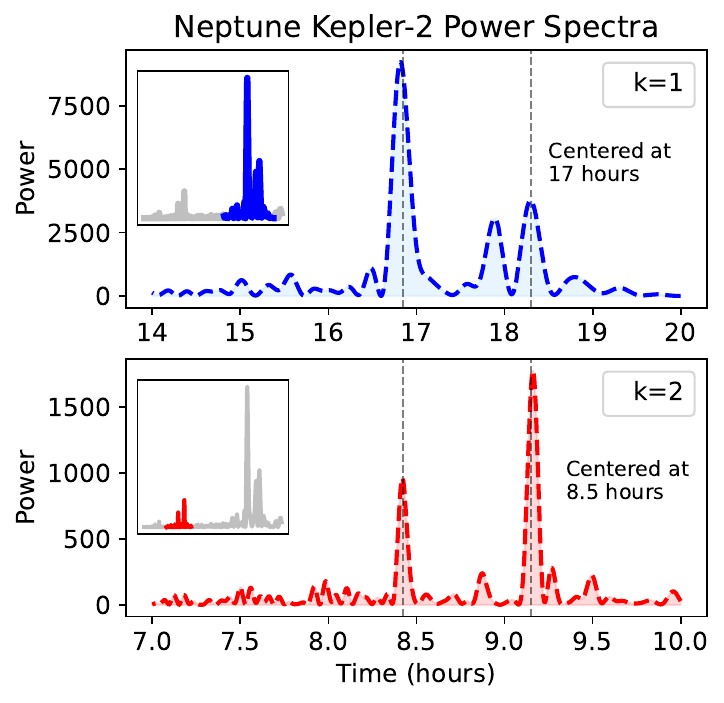}}
\caption{Comparing $k=1$ and $k=2$ wavenumbers on Luhman 16 AB periodograms and Neptune power spectra. {The vertical lines represent the boundary used to calculate the relative range of period structures. Black, dashed lines represent multiple-of-two of the range; i.e.  [4.5, 5.5] hours are 2 times the [2.25, 2.75] hours range. The colored, dashed lines in each plot of Panel \textbf{(a)} represent the multiple-of-two locations of the peaks from the other plot; i.e. the k=1 plot contains location of k=2 peaks multiplied by 2, the k=2 plot contains location of k=1 peaks divided by 2. }
\newline
Panel \textbf{(a)} -- Luhman 16 B data periodograms both contain multi-peaked period components at 4 to 6 hours ($k=1$ wavenumber, blue) and 2 to 3 hours ($k=2$, red). The period multiples do not match: the $k=1$ largest peaks do not correspond to the $k=2$ largest peaks as shown by the dotted lines. Inset plots show the relative powers and positions of two structures. The relative period range is defined as $(P_\text{max}-P_\text{min})/P_\text{min}$.  \\
Panel \textbf{(b)} -- Neptune power spectra from \citet{simon_neptunes_2016} showing the power of each period component arising to rotational modulation of high-altitude cloud bands. There exists matching similarity in the largest peaks at 16--19 hours ($k=1$ waves) and at 8.0--9.5 hours ($k=2$ waves).
\label{fig:periodograms_k1-Andk2}}
\end{figure*}




\subsection{Results for periodogram analysis}\label{subsec:periodogramResult}

\textbf{Broad range of periodicity; component B dominates the variability:} The data itself contains multiple timescales of periodicity corresponding to short and long-period variations. The top panel of Figure \ref{fig:XYposDistance_periodogram} shows that the largest contribution to Luhman 16 AB variability comes from periods under 10 hours and periods above 50 hours. There is little power between periods of 125 and 200 hours in the Luhman 16 AB data.

The strongest powers are found at around 5 hours -- a well-supported rotational period value of Luhman 16 B from numerous past observations \citep[][]{apai_tess_2021, buenzli_cloud_2015, gillon_fast-evolving_2013, burgasser_resolved_2013}.
In the HST analysis by \citet{buenzli_cloud_2015} and TESS analysis by \citet{apai_tess_2021}, a strong power peak around 7.5 hours has been attributed to Luhman 16 A rotational period. 

{However, the rotational period of Luhman 16 A around 7.5 hours in our TESS data is significantly smaller in power relative to the $\sim$5-hour power of Luhman 16 B and is well below the FAP level, as shown in the bottom panel of Figure \ref{fig:periodogram_windowFunction_Data}, and in Figure \ref{fig:periodogramCompare_A_and_B}. The power around Luhman 16 A rotational period is also significantly smaller in our TESS Sector 36 \& 37 data compared to Sector 10 data from \citet{apai_tess_2021}. Figure \ref{fig:periodogramCompare_A_and_B} shows four periodograms for every consecutive 13-day data collection segment that make up the overall 52-day dataset and compares the power amplitude of Luhman 16 A and B variability. The power in the periodogram close to the Luhman 16 A rotation period remains consistently very low -- nearly flat --- through the duration of the TESS observations.}

{The cause for the lower rotational modulation amplitude of Luhman 16 A in our 2021 dataset (compared to the 2019 dataset in \citealt{apai_tess_2021}) is not known. Long-term changes in Luhman 16 A atmosphere could be responsible for the difference in the photometry taken two years apart, as shown by \citet{bedin_hubble_2017}. Due to the very low power around the rotation period of component A, in this study, we attribute most of the binaries' variability to the more dominant component B.}

\textbf{Strong power in window function}: The window function periodogram, similar to the positional variation, also shows significant power for periods from 50 hours to 200 hours, and very little power for periods under 25 hours, as shown in the first and second panels of Figure \ref{fig:periodogram_windowFunction_Data}. The short-period variation is thus significantly less contaminated compared to the longer-period data and could be a safe domain for analysis.

\textbf{Significant long-period power in spacecraft pointing variation}: The spacecraft pointing variation is evaluated using the X position, Y position, and median variation of the position of the target on the detector. 

The second, third, and fourth panels of Figure \ref{fig:XYposDistance_periodogram}, show that the power of these positional variations only exists for periods longer than 25 hours. Some of these power peaks coincide with power peaks from the data periodograms (i.e., 63-hour peak), while others create 'gaps' with no significant power (i.e., 72 and 90-hour peaks). Lastly, there exist very large window function power peaks from 125 to 200 hours not seen in the data.

Periodograms analysis shown in Figures \ref{fig:XYposDistance_periodogram} and \ref{fig:periodogram_windowFunction_Data} show a dataset rich in periodicity ranging from periods under 10 hours to periods as long as 200 hours. At the same time, the spacecraft pointing variations and the window function also show a strong, non-uniform presence across a range of periods, which are potential contaminants in specific parts of the data but not others. Here, the long-period analysis (above 50 hours) would require more careful biases assessment than the short-period (under 10 hours) variability. 

Here, we discuss in detail the analysis of short-period variability in the periodogram of Luhman 16 B.

\textbf{Periodogram fit using synthetic sine waves could explain the primary 5-hour power peak:} 

Using the simple multi-sine model outlined in Section \ref{subsec:periodoSyntheticSineFit}, we generated \textit{periodogram fit} using synthetic light curves with the single-sine, three-sine, and six-sine models and generated GLS periodograms for each. The second, third, and fourth panels of Figure \ref{fig:periodogram_fits} are the periodograms of the corresponding synthetic light curves. 

The three-sine and six-sine models capture the power peaks better than the single-sine model, and more sine waves brought the fits closer to the 1200-hour duration periodogram. The number of peaks between periods of 4--6 hours will decrease with shorter data duration from which the periodogram is generated, showing that the period distribution in the light curve changes and becomes more complex with time. Following \citet{apai_tess_2021}, we started with the three-sine model as a simple approach to fitting the light curve, which works well for segment up to 30-hour in duration. In section \ref{sec:LCfit} we will discuss generating light curve fit in the time domain using a three-sine light curve model with phase offset information.

The primary structure with the strongest power around 5 hours seems to contain multiple sub-structures. The first panel of Figure \ref{fig:periodogram_fits} shows that the data periodogram contains many peaks between 4 and 6 hours. Two maxima are located near 4.7 hours and 5.25 hours respectively. A comparison with the 2019 data from \citet{apai_tess_2021} shows that there are significantly more period peaks in the period range from 4 to 6 hours, due to the three times higher cadence and longer baseline of the new 2021 data set.

\textbf{Power in $k=1$ and $k=2$ wavenumbers:} 

We identify prominent peaks in the periodogram power around 2.5 hours, which is half of the rotational period (5-hour). The strongest periodogram peak is attributed to Luhman 16 B rotational period of around 5 hours \citep[][]{apai_tess_2021, buenzli_cloud_2015}. Following \citep[][]{apai_tess_2021}, we interpret the 2.5-hour peak group as half-periods, corresponding to k=2 wavenumbers, as in standing waves: $\lambda\propto \frac{2L}{n}$ where $n$ is a natural number.

We examined in detail the primary 5-hour power peak and the half-period 2.5-hour peak in Figure \ref{fig:periodograms_k1-Andk2}a. Here onward we refer to the primary peaks around the 4--6 hours range as the $k=1$ periods and the secondary peaks around the 2--3 hours range as the $k=2$ periods. 

The period distributions of $k=1$ and $k=2$ periods both contain dominant double-peak features and multiple smaller peaks. These periods are centered at 2.5 hours and 5 hours respectively, suggesting that they are multiples of each other. 

Figure \ref{fig:periodograms_k1-Andk2}a shows that the $k=2$ periods have a relative range smaller than that of the $k=1$ periods: 10.5\% versus 22.2\%. The relative period range is defined to be $(P_\text{max}-P_\text{min})/P_\text{min}$, where $P_\text{max}$ and $P_\text{min}$ is 4.5--5.5 hours and 2.375--2.625 hours for the $k=1$ and $k=2$ periods, respectively.
Upon closer inspection, the locations of the largest $k=2$ peaks are not multiples of 2 of the largest $k=1$ peaks. For example, multiplying the peaks at 2.43 and 2.575 of $k=2$ periods does not reproduce the location of the peaks at 4.70 and 5.25 hours of $k=1$ periods. 

\begin{figure*}
    \centering
    \includegraphics[width=\textwidth]{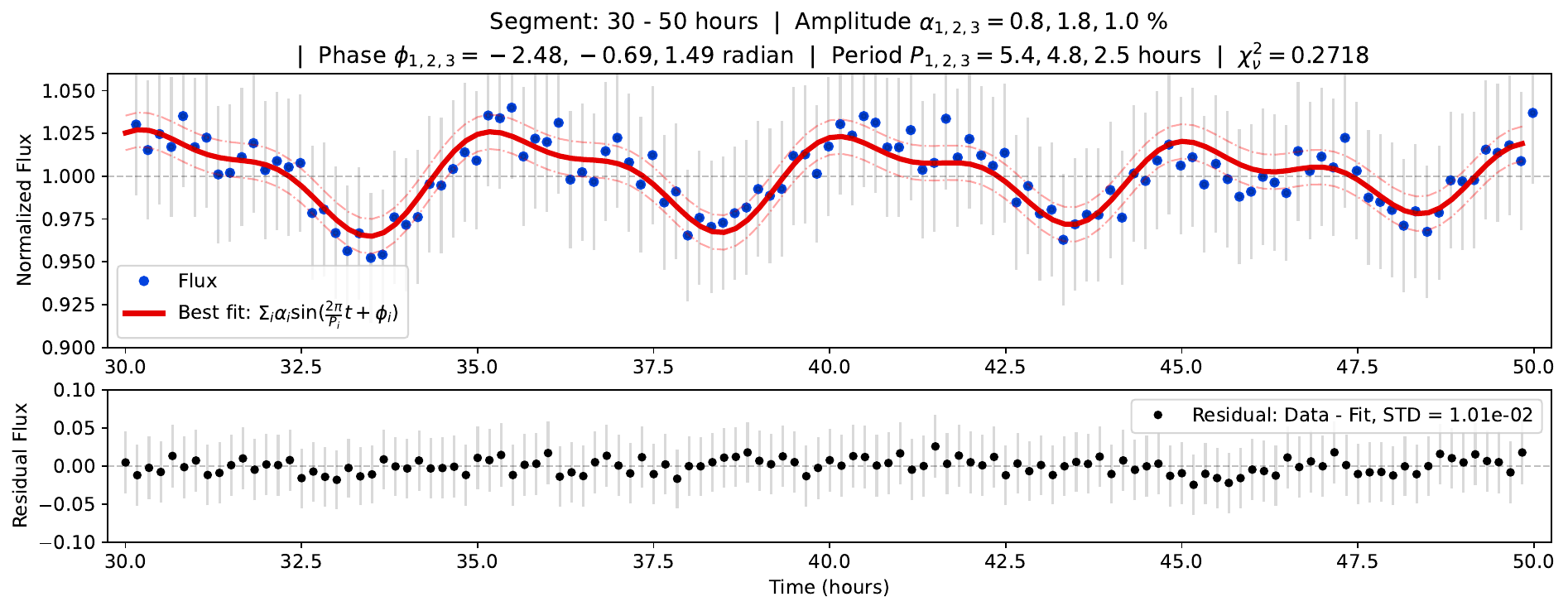}
    \caption{\textbf{\textit{Upper}}: The light curve fit of the 30-50 hours segment using the three-sine model (red curve) corresponding to the data (blue points). Thin vertical lines show the photometric error. Amplitude, period, phase information, \textbf{and the reduced chi-squared ($\chi_\nu^2=0.2718$)} are displayed in the title. 
    \textbf{\textit{Lower}}: the fit residual (data minus fit). The standard deviation (STD) of the fit residual is displayed in the legend.
    \label{fig:MCMC_sector3637_30_50-hr_fit}}
\end{figure*}

It is possible that the half-period $k=2$ peaks represent the higher-order wavenumbers propagating zonally in the atmospheres. {The $k=2$ wavenumbers have been previously observed in Luhman 16 B (\citealt{apai_tess_2021}), in two other L/T transition brown dwarfs \citealt[][]{apai_zones_2017} and also in the power spectrum of Neptune}. Figure \ref{fig:periodograms_k1-Andk2}b shows the Neptune power spectra taken from the \textit{Kepler/K2} \citep[][]{howell_k2_2014} observation of Neptune with a cadence of 30 minutes over 50 days (data from \citealt{simon_neptunes_2016}, power spectra generated by \citealt{apai_zones_2017}). 

\citet{chavez_evolution_2023} compiled infrared observations of Neptune from 1994 through 2022 and determined the cloud activity is variable in an 11-year cycle -- with minima in 2002 and maxima in 2015 when the \textit{Kepler/K2} data is taken -- with strong correlation to solar activity and implication of cloud-top photochemistry by UV radiation from the Sun.

Multi-peaked structures both appeared clearly around the half-period and primary period of Neptune's rotation, i.e. 9 and 18 hours. {However, compared to Luhman 16 B, the $k=1$ and $k=2$ periods in Neptune's power spectra are much more evenly matched - their relative range remained the same. The $k=1$ and $k=2$ periods of the Neptune data are almost perfectly in a one-to-two relationship, in contrast to Luhman 16 B. The exact mechanism to explain this is not well-known, but it is possible that the Neptune variability is mainly due to high-atmosphere cloud-top ices, intensified by solar UV radiation during solar maximum \citep[][]{simon_neptunes_2016} that contributed to the variability observed with \textit{Kepler/K2}. Zonal circulation is highlighted by ice particles as tracers of wind and results in the variability structure over the 49-day observation of Neptune. If deeper zonal circulation lower in the atmosphere had been visible, it would likely introduce more varied windspeed distribution which would not result in perfect one-to-two relationship between $k=1$ and $k=2$ periods as observed in the Neptune data.}

Further analysis of this correspondence will be presented in Section \ref{subsec:discuss_k2wavenumber}, where we will examine Solar System gas giants windspeed distribution as a function of latitude.

\vspace{-5mm}
\begin{deluxetable*}{rccccc}
\tablecaption{Parameters and fit-result for multi-sine $\Sigma_i [\alpha_i \sin (\frac{2\pi}{\omega_i} t + \phi_i )]$ models in Figure \ref{fig:MCMC_sector3637_30_50-hr_fit}, \ref{fig:MCMC_sector3637_183_203-hr_fit}, \ref{fig:MCMC_sector3637_409_433-hr_fit} and \ref{fig:MCMC_4sines_sector3637_408_488-hr_fitpanelB}.}
\tablehead{Segments $t$ & Periods $\omega_i$ (hours) & Amplitudes $\alpha_i$ ($\%$) & Phase $\phi_i$ (radian) & $\sigma$[Residual] ($\%$) & {Reduced chi-squared $\chi^2_\nu$}} 
\startdata
30-50 hr, F\ref{fig:MCMC_sector3637_30_50-hr_fit}     & 5.4, 4.8, 2.5     & 0.8, 1.8, 1.0         & -2.48, -0.69, 1.49 & 1.01 & 0.2718 \\
183-203 hr,  F\ref{fig:MCMC_sector3637_183_203-hr_fit}     & 4.9, 4.9, 2.5     & 2.7, 1.6, 1.9     & -2.82, -0.21, 2.80  & 1.17 & 0.2675 \\
409-433 hr, F\ref{fig:MCMC_sector3637_409_433-hr_fit}     & 4.5, 4.6, 2.4   & 0.4, 1.8, 1.1       & 0.39, 2.72, 1.90  & 1.21 & 0.4240 \\
410-490 hr, F\ref{fig:MCMC_4sines_sector3637_408_488-hr_fitpanelB}  & 2.6, 2.4, 5.4, 4.8    & 0.5, 0.9, 1.1, 2.0        & 1.18, 0.0, 14.70, 9.34     & 1.21 & 0.7412
\enddata
\end{deluxetable*}

\section{Short-Period Light Curve Fits} \label{sec:LCfit}

In the following analysis,  we refer to "short periods" when speaking about variability with periods shorter than 10 hours. The previous periodogram analysis shows that 1) injecting multiple sine waves with periods close to the rotational period of the object could explain the periodogram, and 2) there exists multiple short and long-period variations in the data. The intensity variations on timescales shorter and longer than rotational periods are likely to have different physical origins; this is a motivation to examine them separately. In the following section, we will explore the data for short periods on par with the rotational period and attempt to fit the short periods with a light curve model composed of planetary-scale waves.

\subsection{Fitting the Short-period light curve with a multi-sine model}

 To capture only the short-period peaks, a two-step process is employed: filtering the peaks with periods longer than 20 hours with a box-car average smoothing algorithm, and then subtracting the long-period light curve from the original data. 

We selected three segments in the light curve for light curve fitting. We identified three different segments about 20 to 30 hours in duration. These segments correspond to 30-50, 183-203, and 409-433 hours, counting from the beginning of data collection on BTJ day 2280.30 in Sector 36. We also selected a longer segment at 410-490 hours with a duration of 80 hours.

For the 20 to 30 hours segments, we fitted the normalized light curve with a model comprising three sine waves in the form:
\begin{align}
    1+\Sigma^3_i a_i  \sin (\omega_i t + \phi_i) 
\end{align}

with a total of 9 parameters for fitting, following \citet{apai_tess_2021}. 

The three-sine model is the simplest initial model to explore the validity of planetary-scale waves. For the 80-hour segment, we fitted the light curve with a four-sine model following the same format with a total of 12 parameters. Only in the case of increased data quantity, such as with the 80-hour window did we include an additional sine wave to improve fit quality.

The fitting process is comprised of two steps: 1) finding an initial guess and 2) MCMC (Markov-Chain Monte Carlo) fitting. Firstly, we find initial guesses for the MCMC model from a fast decision tree optimizer \verb|hyperOPT| (\citealt{bergstra_hyperopt_2015}). We found that fitting methods based on gradient descent do not work optimally with the sinusoidal model. Due to the oscillatory nature of sinusoids, these fits tend to converge at a local minimum. The decision tree optimizer mitigates this problem by exploring the parameter space simultaneously to find a better parameter set that will minimize the difference between the model and data. This step ensures the best initial guesses are obtained, but also takes less time to find best-fit parameters. 

Secondly, we used MCMC to sample and explore the parameter space with \verb|emcee| (\citealt{foreman-mackey_emcee_2013}). From the initial guess given in the previous step, 'random walks' or random variations in parameters are taken to explore the parameter space and find the best fit. Weights can be given to each parameter to decide their 'random walk' ranges - well-constrained parameters need not deviate too much while less-constrained parameters could benefit from more varied guesses.

\begin{figure*}[ht!]
\centering
\includegraphics[width=\textwidth]{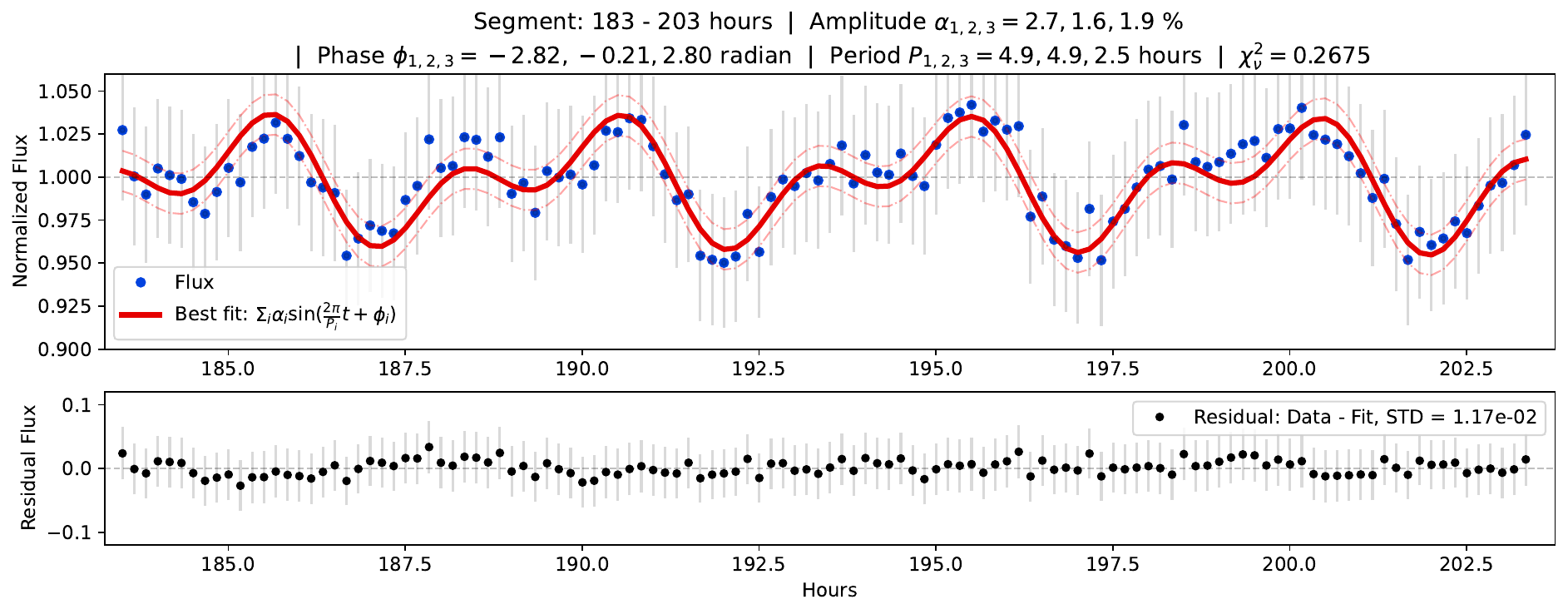}
\caption{\textbf{\textit{Upper}}: The light curve fit of the 183-203 hours segment using the three-sine model (red curve) corresponding to the data (blue points). Thin vertical lines show the photometric error. Amplitude, period, phase information, {and the reduced chi-squared ($\chi_\nu^2=0.2675$)} are displayed in the title. 
\textbf{\textit{Lower}}: the fit residual (data minus fit). The standard deviation (STD) of the fit residual is displayed in the legend. 
\label{fig:MCMC_sector3637_183_203-hr_fit}}
\end{figure*}

\begin{figure*}[htp!]
\centering
\includegraphics[width=\textwidth]{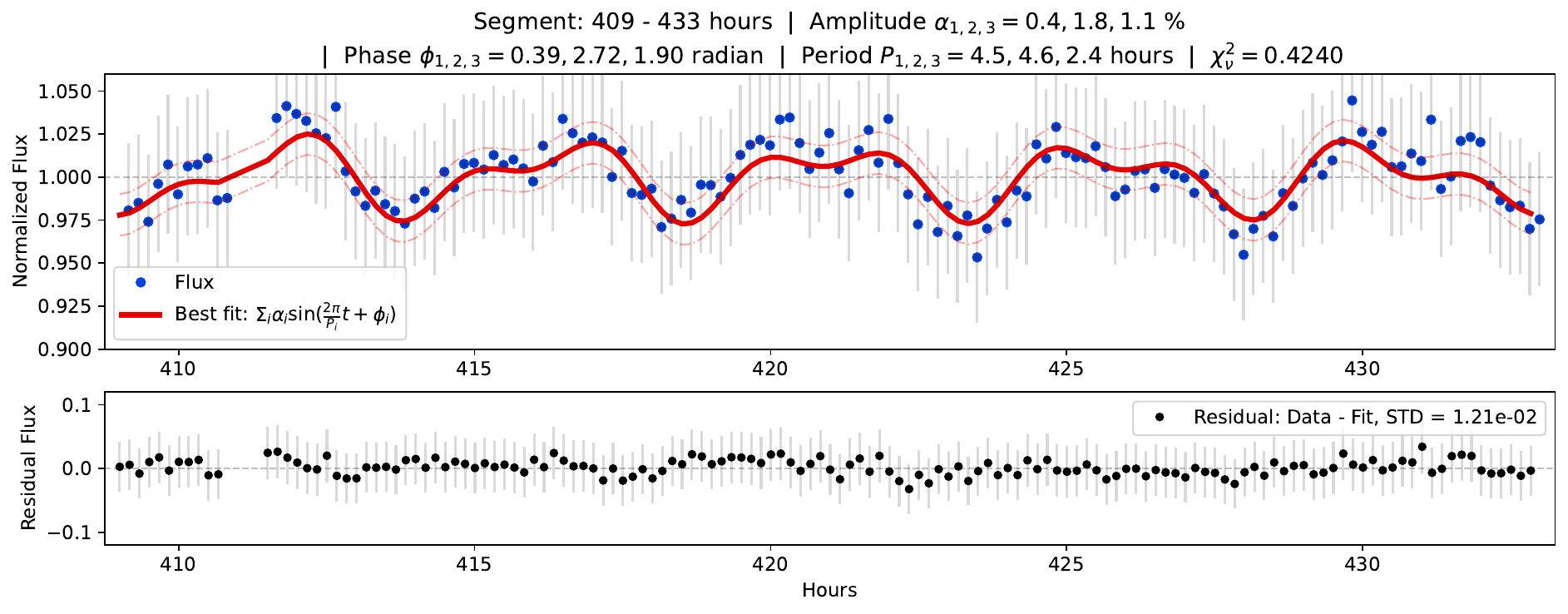}
\caption{\textbf{\textit{Upper}}: The light curve fit of the 409-433 hours segment using the three-sine model fit (red curve) corresponding to the data (blue points). Thin vertical lines show the photometric error. Amplitude, period, phase information, and {the reduced chi-squared ($\chi_\nu^2=0.4240$)} are displayed in the title. 
\textbf{\textit{Lower}}: the fit residual (data minus fit). The standard deviation (STD) of the fit residual is displayed in the legend.
\label{fig:MCMC_sector3637_409_433-hr_fit}}
\end{figure*}
\subsection{Results from Light-curve Fitting}\label{subsec:LCFitresult}

\textbf{Three-sine model fits the light curve well for the duration of $\sim$30 hours:} We show in Section \ref{subsec:periodogramResult} that the light curve could be fitted using combinations of period components as a multi-sine model in the frequency space. To demonstrate this in the temporal space, we created test segments with a duration of 20 to 30 hours and fitted them with the multi-sine model. The resulting light-curve fits for the 30-50 hour, 183-203 hour and 409-433 hour segments are shown in Figures \ref{fig:MCMC_sector3637_30_50-hr_fit}, \ref{fig:MCMC_sector3637_183_203-hr_fit}, and \ref{fig:MCMC_sector3637_409_433-hr_fit}, respectively. All three segments are fitted with three-sine wave models. The fit residuals are also shown within the bottom panel of each plot.

Results of the light curve fitting for the first segment within 30-50 hours are shown in Figure \ref{fig:MCMC_sector3637_30_50-hr_fit}. The three-sines model shows very good correspondence with the light curve data, with two periods around 5 hours and one period around 2.5 hours that best match the data. This period distribution reflects the periodogram analysis in Section \ref{subsec:periodogramResult}, where the highest-power peaks for periods under 20 hours both correspond to the $k=1$ and $k=2$ waves, respectively. 

Our analysis of the other two segments showed similarly consistent results. Figure \ref{fig:MCMC_sector3637_183_203-hr_fit} and Figure \ref{fig:MCMC_sector3637_409_433-hr_fit} show the light curve fits for the 180-210 hours segment and the 409-433 hours segment, respectively. Particularly, the best-fit periods correspond well to the peaks in the periodograms that are identified as $k=1$ and $k=2$ waves in Section \ref{subsec:periodogramResult}. 

We calculate the standard deviation of the fit residual (data minus fit) to estimate the goodness of the model. The respective standard deviation of fit residual for each model in Figure \ref{fig:MCMC_sector3637_30_50-hr_fit}, \ref{fig:MCMC_sector3637_183_203-hr_fit}, \ref{fig:MCMC_sector3637_409_433-hr_fit} are 1.02\%, 1.17\%, 1.14\% respectively. These small residual shows that the sine-wave models well-captured all the components present in our light curve data on a timescale of 20-30 hours. The fit residuals for the light curve fit with our sine wave model are consistent with past results of \citet{apai_tess_2021}, which also found period components about 2.5 and 5 hours via the multi-sine model to fit the light curve best. Nominal photometric noises in the data are similar ($\approx4\%$), and standard deviations of the fit residual are also similar ($\approx1\%$).

\textbf{Parameters in fit show small variation from one test segment to another:} Throughout each segment shown in our analysis, the amplitudes in the multi-sine model are all non-constant as we move from one segment to another. {This suggests that the amplitudes of period components experience a time evolution, pointing to a dynamic picture in the data. Moreover, the phase offset parameters also experience changes throughout the segment fits although these variations seem to be within a given bound. The phase parameters $\phi_i$ (in radians) for each segment in Figure \ref{fig:MCMC_sector3637_30_50-hr_fit}, \ref{fig:MCMC_sector3637_183_203-hr_fit}, \ref{fig:MCMC_sector3637_409_433-hr_fit} are: ($-2.48$, $-0.69$, 1.49), and ($-2.81$, $-0.21$, 2.80), and (0.39, 1.90, 2.72).}
\newline

\textbf{Four-sine model fits the light curve well for the duration of 80 hours:} 

\textbf{From Figure \ref{fig:XYposDistance_periodogram} and \ref{fig:periodogram_windowFunction_Data}, the periodogram shows} that the period distribution becomes more complex as one considers a longer duration of light curve evolution, which points to a non-static, actively evolving picture of the atmosphere. In the 410-490 hours segment - the longest segment we have in this analysis - we used a four-sine model to try to fit the light curve as having one more sine wave helps produce better fits. 

\begin{figure*}[htp!]
\centering

\includegraphics[width=\textwidth]{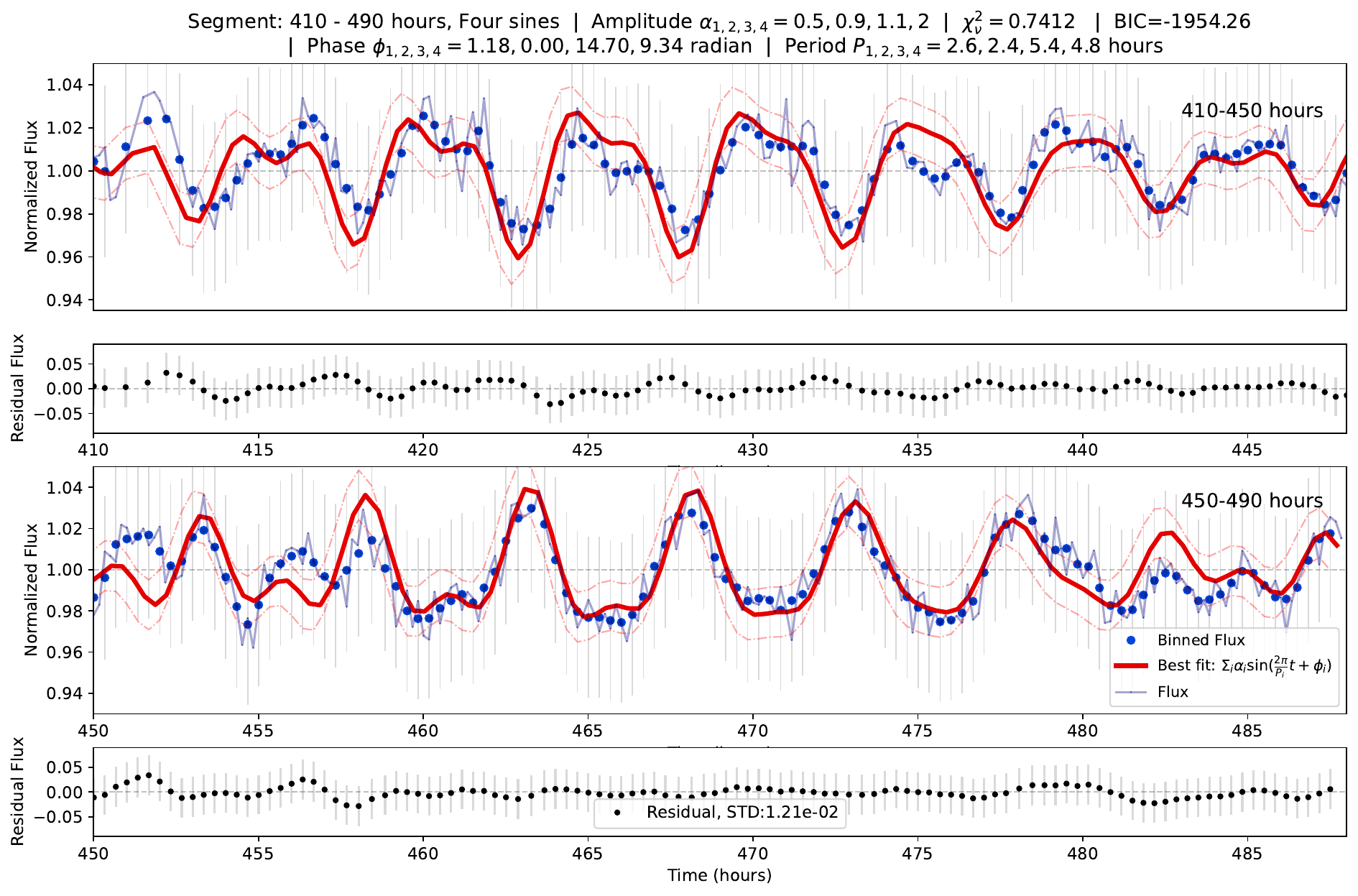}
\caption{Light curve fit for the 410-490 hours segment. The light curve fit uses a four-sine model (red curve) on the binned data with a 20-minute cadence (blue points), based on the 10-minute cadence original flux (thin blue line). Thin vertical lines show the photometric error. {Amplitude, period, phase information, the reduced chi-squared ($\chi_\nu=0.7412$), and the Bayesian Information Criterion score (BIC=$-1954.26$) are displayed in the title (for a discussion of BIC interpretation, see Appendix \ref{appendix:3sine_bic}). The fit residuals (data minus fit) and the standard deviation (STD) of the fit residuals are also shown. For the result of a three-sine fit for the same data segment, see Figure \ref{fig:hyperOPT_3sines_sector3637_408_488-hr}.}
\label{fig:MCMC_4sines_sector3637_408_488-hr_fitpanelB}}

\end{figure*}

Moving from a 20-hour to an 80-hour window, the amount of photometric data quadrupled for this test segment compared to the rest. We binned the light curve down from a 10-minute cadence to a 20-minute cadence after having applied a box-car median filter. The result for the 80-hour fit is shown in Figure \ref{fig:MCMC_4sines_sector3637_408_488-hr_fitpanelB}. 

The four-sine model fitted the light curve well and resulted in a 1.14\% standard deviation of the residual, with a reduced chi-square $\chi_\nu^2 \sim 0.74$. Notably, the solution converged on a range of period components very similar to the primary and secondary power peaks in the periodogram: 2.4, 2.6 hours ($k=2$ waves), and 4.8 and 5.4 hours ($k=1$ waves). This correspondence is consistent with our analysis in Section \ref{sec:Periodo}, and also with the periods found for shorter segments with the three-sine model. The residual of the 80-hour fit is comparatively more varied than the shorter segment fits, hinting that remnants of periodic structures seem to persist within the fit. 
 
{We find that the four-sine model matched well the lightcurve behavior from 410-490 hours, better than the three-sine model. In Appendix \ref{appendix:3sine_bic}, we discuss the quality difference between the three-sine model and the four-sine model fit for this 80-hour data segment. We also used a Bayesian Information Criterion score (BIC) \citet{bauldry_structural_2015} to quantify the goodness of fit as well as over-fitting risk. Through that analysis, we found that the four-sine model produced a better fit compared to the three-sine model while remaining a simple enough model to explain the light curve behavior for this particular data segment.}



\section{Long-period light curve evolution}\label{sec:LongPeriodLCevolution}

Luhman 16 AB has been shown to exhibit variable behavior on timescales longer than 20 hours (\citealt{apai_tess_2021}). In this section, we present the analysis of long-period variability components on Luhman 16 AB tens of times the rotational period. Long-period variations in the light curve are responsible for brightness variability of about $\pm5\%$ in the light curve, similar to the short-period variation. In the following analysis,  we refer to "long periods" when discussing variability with periods longer than 15 up to 100 hours.

\subsection{Frequency filtering of long-period variations}

Before we proceeded with the analysis, we assessed to understand the potential contamination of background sources to the long-period data. We obtained a list of sources within the vicinity of 2 arc-minutes of Luhman 16 along with the periodogram of their PATHOS-extracted light curves. We find that the sources bright and variable enough to present significant contamination comparable to Luhman 16 AB are all separated by more than 200 arcseconds (about 10 pixels). Sources under 200 arcseconds separation otherwise have featureless periodograms and no significant variability (see Appendix). Looking at the periodograms of the background sources' light curves, no variability is found in the long periods above 40 hours to under 100 hours, and the periodogram appears featureless. In agreement with the analysis conducted with background sources using HST from \citet{apai_tess_2021}, this indicates that background source contamination in the vicinity of Luhman 16AB is of minor concern for long-period data under 100 hours. 

The periodogram analysis in the first panel of Figure \ref{fig:periodogram_windowFunction_Data} shows that the long periods from 50 to 125 hours contain the highest amplitude power, comparable to the amplitude of the short periods around 5 hours. However, the long period range coincides with strong contamination from the window function, as evident in the second panel of Figure \ref{fig:periodogram_windowFunction_Data}. 

Considering the window function contamination, there is likely a significant spurious signal in the periodogram above 150 hours and relatively significant power contamination for periods between 50 to 125 hours. However, there are `gaps' of periods retrievable in the light curve data, and this range is also where the data shows the strongest power in long-period variability. 

For potential lightcurve signals from telescope positional drift, the same conclusion can be drawn for periods above 150 hours in Figure \ref{fig:XYposDistance_periodogram}: There is a strong possibility for periods contamination at this range.

Thus, we considered only the 15-95 hours range for our long-period analysis since this range contains the highest power with relatively less contamination coming from the windowing function and potential pointing errors.

\begin{figure*}
    \includegraphics[width=\textwidth]{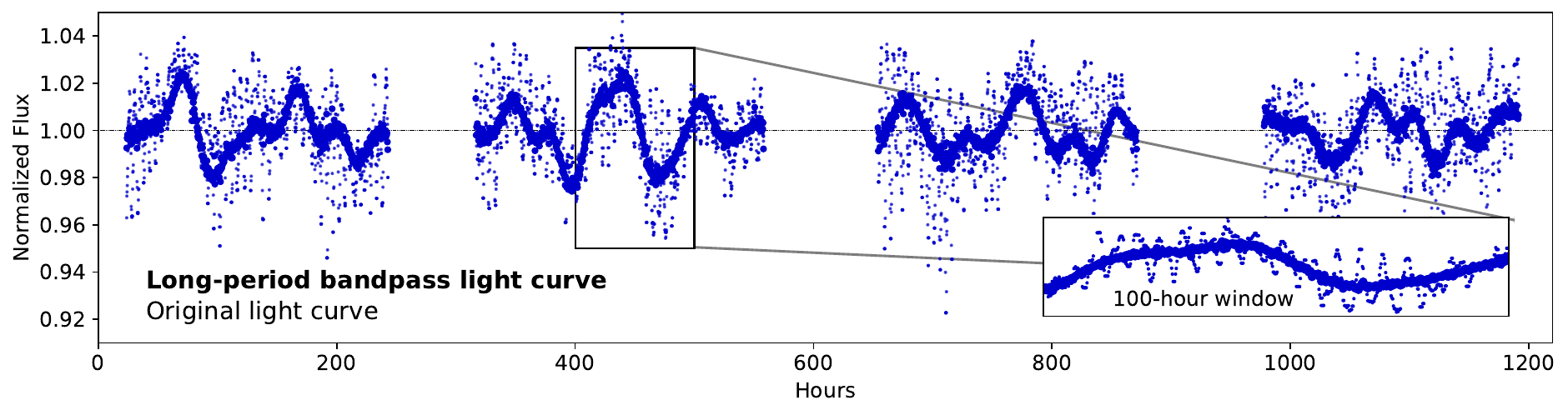}
    \centering
    \caption{Resulting light curve from filtering only the periods within 15-95 hours using a Fourier bandpass (bold points). The background data points display the original light curve data. The inset plot shows that the long-period filter captures well the mean evolution of the data. \label{fig:Longperiod_bandpassFilterLC}}

    \includegraphics[width=\textwidth]{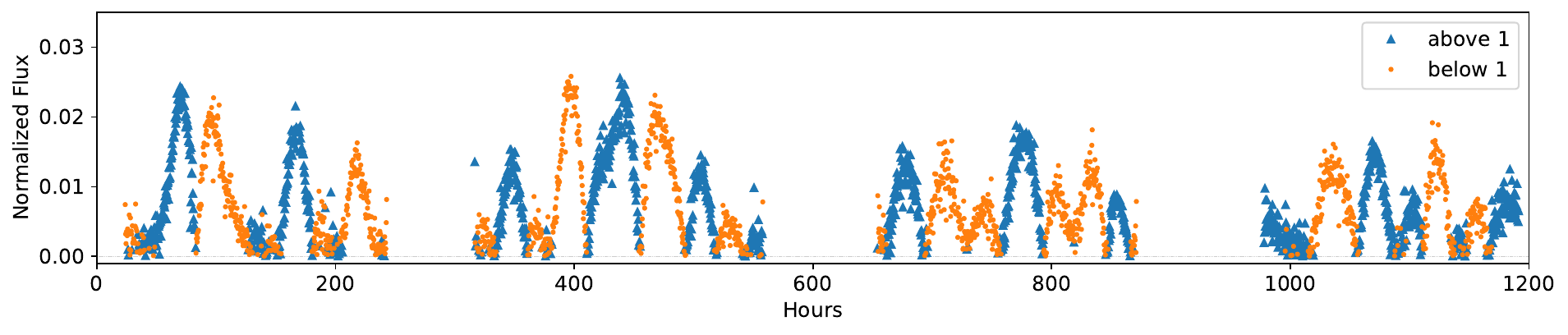}
    \centering
    \caption{Comparing the light curve below and above the 'equilibrium' value of 1 by mirroring the flux value smaller than the equilibrium. If the flux variation is not balanced across the equilibrium baseline, it could be an indication of excess flux arising from a physical feature in the atmosphere (e.g. storm spot).\label{fig:LongPeriodLC_flippingCompare}}
\end{figure*}

In order to filter out periods within 15-95 hours, we employed a Fourier bandpass filter process. First, we interpolated the processed light curve data to create a uniformly-space light curve. Then, we applied a Fourier transform to the light curve to the frequency space and applied a bandpass such that values outside the target period range are zero. Lastly, we applied an inverse-Fourier transform to obtain a time-domain, long-period light curve. The resulting bandpass-filtered light curve is shown in Figure \ref{fig:Longperiod_bandpassFilterLC}, where the long-period light curve is plotted over the original data, demonstrating that this long-period extraction captures the mean of the data very well.

\subsection{Analysis Results of the Long-period Light Curve}\label{subsec:LongPeriodresult}

The evolution of the Luhman 16 AB light curve is markedly different over long periods from its behavior over short periods. We tried fitting the multi-sine wave model and found that the model could not explain the long-period light curve in any segment longer than 100 hours, hinting that the long-period light curve is not time-stationary (i.e. a time-stationary series has statistical properties (e.g., mean and variance) that do not vary in time: for example, a sine wave with constant amplitude, phase and period). As shown in Figure \ref{fig:Longperiod_bandpassFilterLC}, the patterns of flux variability are in the shape of sharp peaks that evolve sporadically in a seemingly non-stationary manner.


First, we compare the flux values below and above the equilibrium level of 1. By assuming a baseline level of flux and evaluating flux increase and decrease around the baseline, we can identify excess flux that indicates potential atmospheric features, for example, storm spot that produces excess flux on a timescale close to self-rotation. 

Figure \ref{fig:LongPeriodLC_flippingCompare} is the comparison of flux value above and below the equilibrium value of 1, in which flux below 1 is vertically mirrored. Numerically integrating the flux, we find the values 5.4 and 4.8 for the flux below 1 and flux above 1, respectively. Figure \ref{fig:LongPeriodLC_flippingCompare} shows that a dip is usually followed by a peak in flux such that the light curve appears relatively even across the baseline level.

\begin{figure*}
    \centering
    \includegraphics[width=0.98\textwidth]{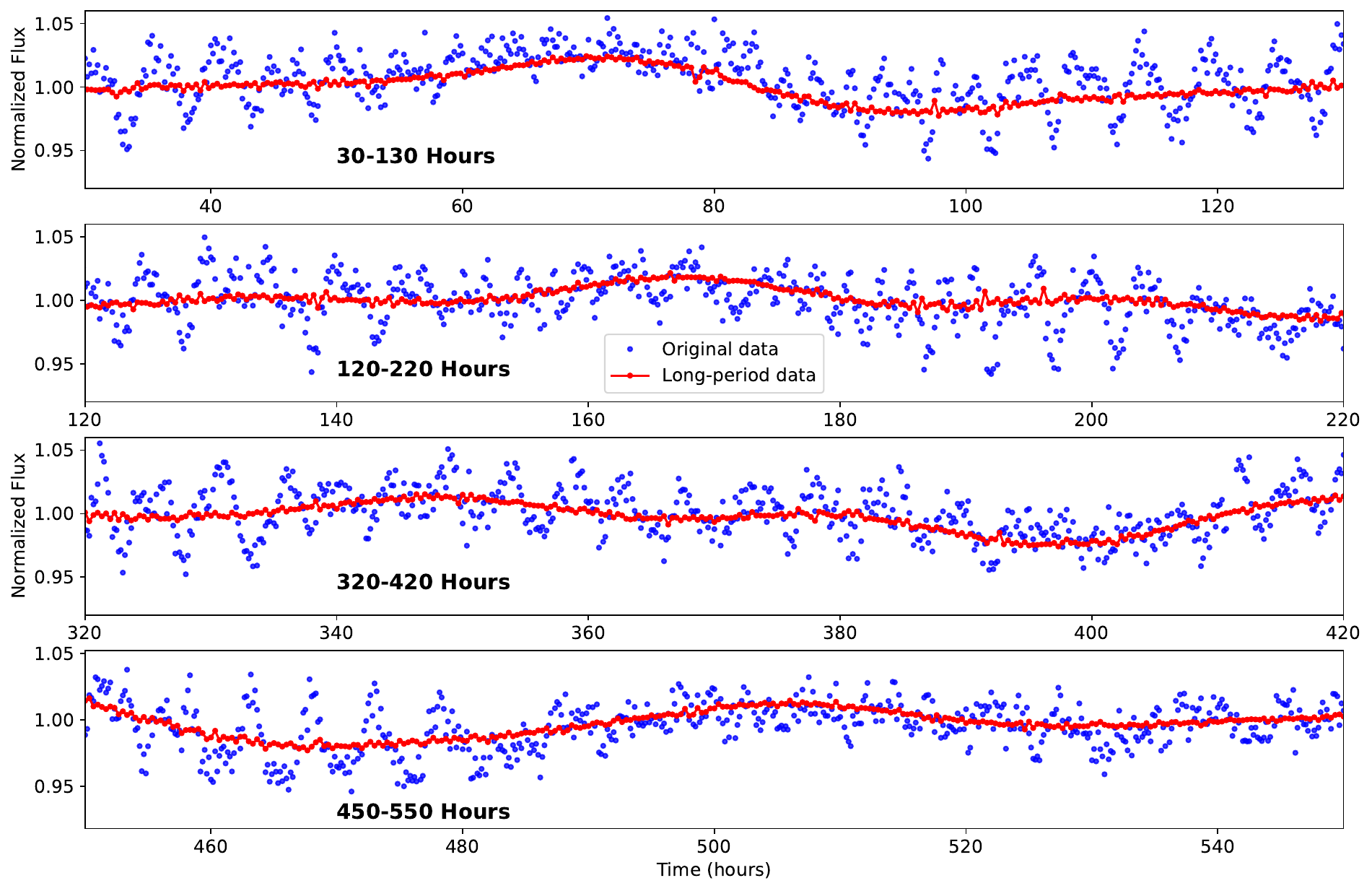}
    \caption{Long-period light curve (red) as a result from frequency filtering of periods component from 15--95 hours. Top-to-bottom panels show different 100-hour segments and the comparison between the original and long-period light curves. 
    \label{fig:zoomOut_longLightCurve}}
\end{figure*}

Figure \ref{fig:zoomOut_longLightCurve} shows a number of shorter segments demonstrating that the long-period filter captures the mean of the data very well. We note a strong correspondence in the long-period evolution with GCM (general circulation model) results from \citet{tan_atmospheric_2021} where different inclined viewing angles for a 3D atmosphere produce different short and long-period evolution.



These timescales of variation are markedly very different from the short-period modulation characterized in the previous sections, hinting at their different physical origin. With this data set, we are demonstrating for the first time that Luhman 16 AB is continuously variable in periods up to 95 hours over a baseline of 1200 hours. In Section \ref{subsec:discuss_longPeriodLC} and Section \ref{subsec:discuss_GCM} we will discuss the potential origins of long-period atmospheric evolution and relevant results from GCM studies.


\section{Discussion}\label{sec:Discussion}

In this section, we discuss the results obtained from the periodogram analysis showing strong power around the rotational period of Luhman 16 B (Section \ref{subsec:periodogramResult}) and characteristics of the narrower range of the $k=2$ waves with respect to the $k=1$ waves. We show via a toy model of Jupiter and Saturn that the narrowing of the period range might be related to the different windspeed distribution between the equatorial region and the mid-to-high latitudes region. Then, we discuss and review possible mechanisms in the literature for the long-period light curve observed in Luhman 16 AB (Section \ref{subsec:LongPeriodresult}).

\subsection{Periodograms}\label{subsec:discuss_periodogram}

Our periodogram analysis of Luhman 16~AB in Sectors 36 \& 37 of TESS monitoring over a time span of 1,200 hours -- covering more than two hundred rotations of the targets -- identified strong peaks around the rotational period of Luhman 16B. In the range between 4.7 to 5.3 hours in our periodogram, there is strong evidence for a rich, multi-peaked distribution of period components (see Figure \ref{fig:periodogram_windowFunction_Data} and \ref{fig:periodograms_k1-Andk2}). Via ou\cite{}r analysis  in Sections \ref{sec:TESSData} \& \ref{subsec:periodogramResult} of spacecraft pointing drift and temporal sampling bias (modeled via a window function), we conclude that periods under 10 hours are unlikely to be contaminated, and emerge from Luhman 16B.  This result agrees with the findings of \citet{apai_tess_2021}, in which variability at the same periods was also attributed to Luhman 16 B. Moreover, as in \citet{apai_tess_2021}, our study also supports the applicability of using the planetary-scale wave model as fits to the periodogram. 


Sectors 36 and 37 TESS data have a cadence of 10 minutes, which is 3 times higher than the 30-minute cadence data used in \citet{apai_tess_2021}. The higher cadence, in combination with the longer baseline, reveals more sub-structures in the periodogram of Luhman 16 B. A multi-peaked distribution of periods with peaks ranging from 4.50 to 5.50 hours is identified (Figure~\ref{fig:periodograms_k1-Andk2}). This period distribution is an interesting correspondence to the Solar System's giant planets, for example, Jupiter (Figure 15 of \citealt{apai_tess_2021}) and Neptune (Figure 3B of \citealt{apai_zones_2017}), which both have multi-peaked power spectra around the rotation periods. 

\subsection{Persistent Period Distribution}

Our new Sector 36 and 37 data show a periodogram for \luB{} that is consistent in all of its properties with the earlier datasets from 2019, presented in \citet[][]{apai_tess_2021}. This is an important finding as it demonstrates that the periodicities present in the targets are sustained over more than three thousand of rotations. 
Although past data are taken prior to TESS \citep[e.g.,][]{gillon_fast-evolving_2013,buenzli_cloud_2015,karalidi_maps_2016}, only provide snapshots in time rather than multi-period light curves, we note that the qualitative behavior of \luB{} described in those studies is consistent with those found here. Thus, there is no evidence for a change in the nature of the rotational modulation and the evidence available is consistent with \luB{} displaying rotational modulations of the same nature since its discovery in 2013 \citep[][]{luhman_discovery_2013}, over the past 10 years.

Therefore, we conclude that the atmospheric processes that modulate the brightness distribution in the atmosphere of \luB{} remained active and dominant over the timescales of at least thousands of rotations (over the course of the available TESS observations between 2019 and 2022), and likely well over 10,000 rotations (since 2013). This persistent nature of the modulations demonstrates that the nature of atmospheric brightness distribution is not due to a transient state, such as a chance cloud alignment, but tied to a fundamental, powerful mechanism within the atmosphere that does not change significantly even over long timelines ($>$10 years). The most obvious source of such a mechanism is atmospheric circulation, which we will explore in the following section.


\subsection{k=2 Wavenumber}\label{subsec:discuss_k2wavenumber}

\begin{figure*}
    \centering
    \includegraphics[width=0.99\textwidth]{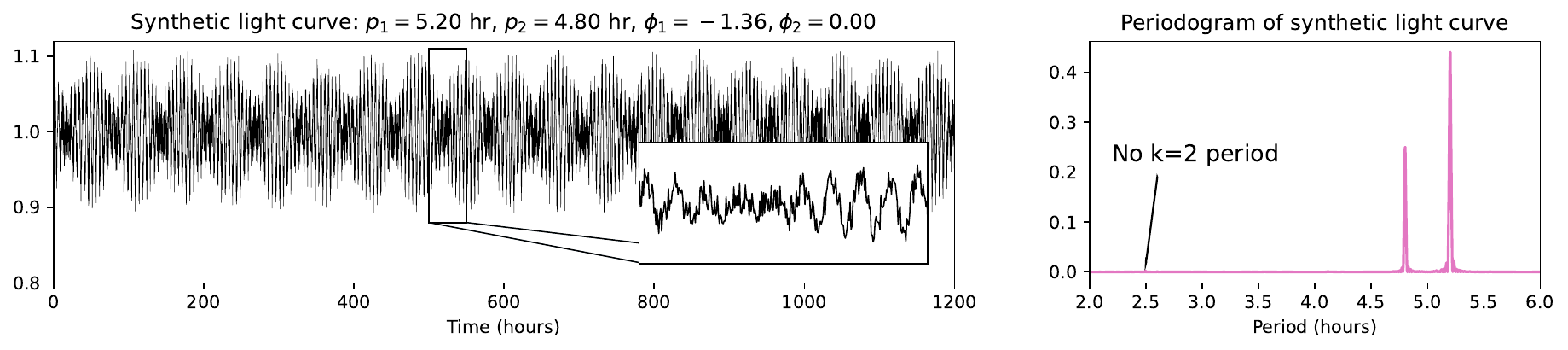}
    \caption{\textit{\textbf{Left}}: Synthetic light curve composed of 2 sine waves with roughly the same periods (5 hours) with added uniform noise. \textit{\textbf{Right}}: The periodogram of this light curve. The synthetic light curve does not show the presence of $k=2$ waves despite having zonally off-phased waves, suggesting $k=2$ period components must come from actual waves in the atmosphere.
    \label{fig:synthetic_zonally_shifted_lightcurve_no-K2_wave}}
\end{figure*}


In this work, we confirmed the existence of group of periods half of the rotational period of \luB{} in the periodograms, which we identify as $k=2$ wavenumber  waves. We find strong similarity between the periodogram power structure of the primary $k=1$ period group and the $k=2$ period group: They are similar in their multi-peaked nature and approximately bimodal distributions of periodogram power (see Figure \ref{fig:periodograms_k1-Andk2}). The presence of $k=2$ periods in our data is consistent with results from two past studies: In \citet[]{apai_zones_2017}, a periodicity analysis of Spitzer light curves of two L/T transition brown dwarfs showed evidence for peaks at half the rotational period. The TESS light curves of Luhman 16~B, presented in \citet{apai_tess_2021}, displayed strong peaks consistent with $k=2$ waves in the generalized Lomb-Scargle periodogram for Sector 10 TESS data. 

The light curves of Solar System gas giants, such as Jupiter and Neptune, all display strong power around the rotational period. For example, the effective period distribution of winds in Jupiter reveals differential rotation (\citealt{porco_cassini_2003}, \citealt{fletcher_how_2020}). \citealt{apai_zones_2017-1} demonstrated that periodogram peaks at half the rotational period also exist in the power spectrum of the Neptune \citep[][]{simon_neptunes_2016} in ways similar to their brown dwarf counterparts. {The general circulation models (GCMs) of Jupiter and Saturn also display prominent banded structures \citep[][]{schneider_formation_2009, lian_generation_2010, young_simulating_2019, spiga_global_2020}, similar to banded structures seen in brown dwarf GCMs \citep[][]{showman_atmospheric_2019, tan_jet_2022}.}

\citet{apai_zones_2017} showed that the peaks around the half rotational period are well fit via the planetary-scale waves model: $k=2$ waves with periods of half of that of the primary $k=1$ waves. The planetary-scale waves modulate the brightness of zones and belts which, in turn, are rotating at different rotational periods due to a combination of zonal circulation and differential rotation. The planetary-scale waves could modulate the thickness and/or structure of condensate clouds \citep[e.g.,][]{apai_hst_2013,vos_let_2022}. 



\begin{figure*}[htp!]
\centering
\subfigure[Cassini Windspeed Measurements of Jupiter from \citet{porco_cassini_2003}]{
    \centering
    \includegraphics[width=0.32\textwidth]{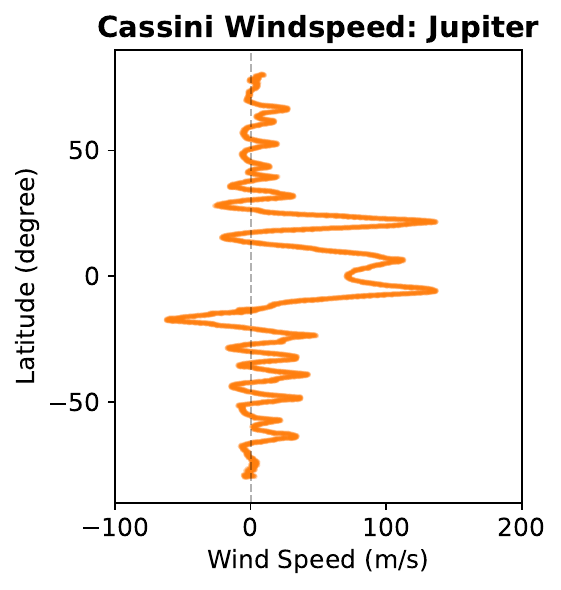}}
\subfigure[Superimposed $k=1$ and $k=2$ periodograms of Luhman 16 AB]{
    \centering
    \includegraphics[width=0.62\textwidth]{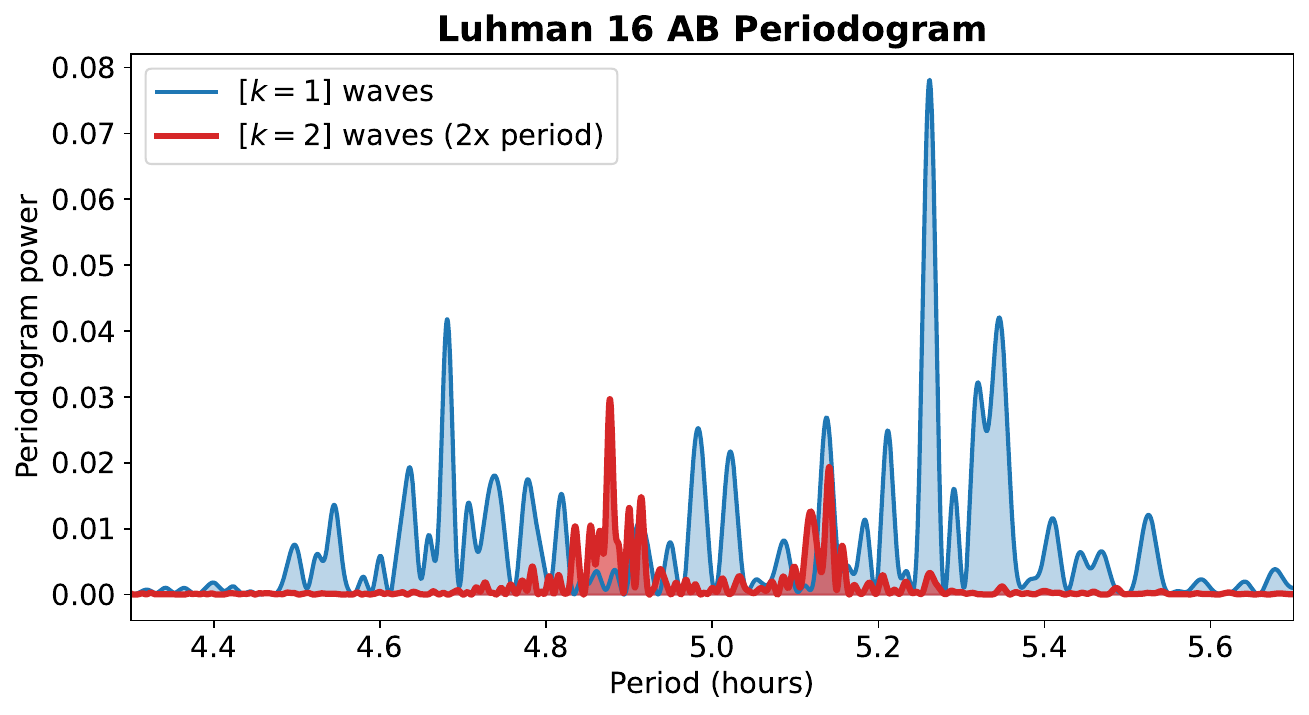}}
    
\quad     
\subfigure[Synthetic Jupiter periods maps and the resulting histograms for two case: full-disks including equatorial and mid-latitudes, and only mid-latitudes to polar region.]{
    \centering
    \includegraphics[width=0.99\textwidth]{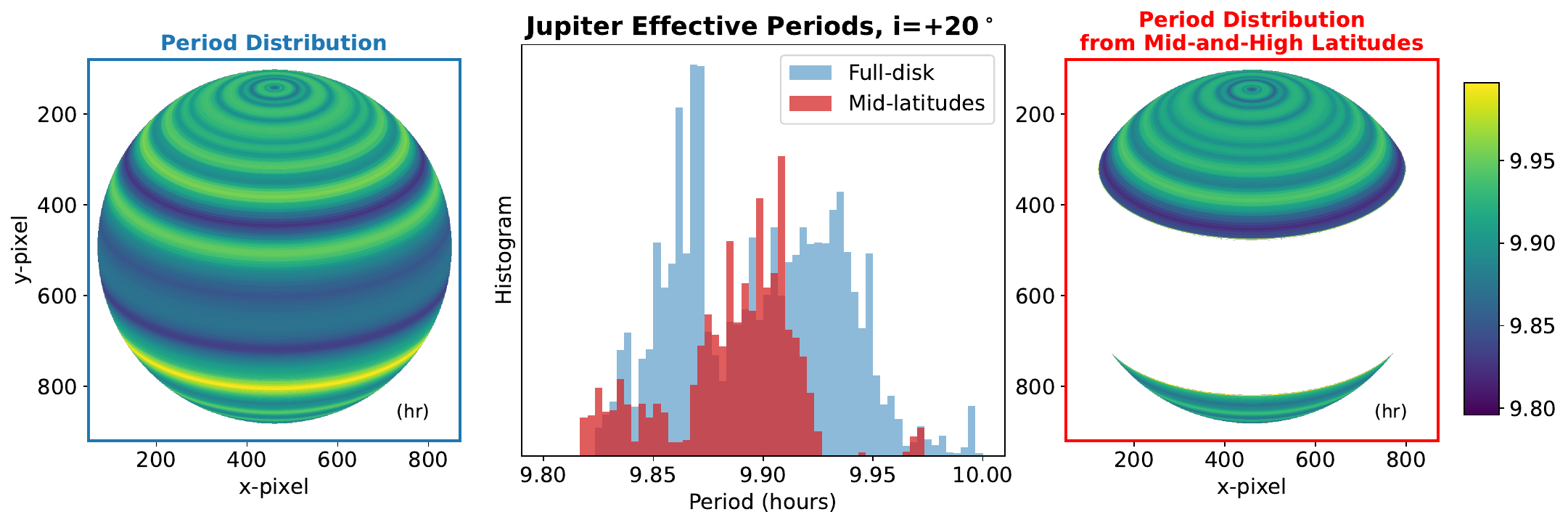}}
\caption{Panel \textbf{(a)} -- The synthetic period distribution of Jupiter based on Cassini windspeed data as a function of latitude from \citet{porco_cassini_2003}. \newline
Panel \textbf{(b)} -- The superimposed $k=1$ and $k=2$ periodograms of Luhman 16 AB, where periods of the $k=2$ waves are scaled up two times such that the 2--3 hour range matches the 4--6 hour range. The $k=2$ waves show a narrower period distribution compared to $k=1$. \newline
Panel \textbf{(c)} -- \textit{\textbf{Left}}: Synthetic image of $20^\circ$-inclined Jupiter showing effective periods distribution of all latitudes; \textit{\textbf{Right}:} Periods distribution only from latitudes above 35$^\circ$N and below 35$^\circ$S; \textit{\textbf{Center}}: Histograms of Jupiter effective period for the two cases. The period distribution significantly narrows in range once the equatorial windspeeds are removed since these latitudes contain the slowest and fastest winds.
\label{fig:CassiniJupiter_model}}

\end{figure*}

In the general circulation models (GCMs) of \citet{tan_atmospheric_2021} with periods of 2.5, 5, and 20 hours, the synthetic periodogram displays large band structures in the equator as a result of strong circulation. It is notable that these models also contained $k=2$ periods. Weak drag seemed to be an important precursor for the formation of large-scale jets: in \citet{tan_jet_2022}, the weak-drag atmosphere model (characteristic drag timescale $\tau_\text{drag}=10^7 s$) contains much larger $k=2$ period power compared to strong-drag atmospheres ($\tau_\text{drag}=10^5 s$). \citet{tan_jet_2022} postulated that two planetary-scale waves with the same zonal frequencies but different zonal phases could result in an apparent feature of zonal wavenumber 2 within the power spectrum. 

However, via a synthetic light curve under the assumption of planetary-scale waves, we show that a $k=2$ period component must correspond to an actual wave in the atmosphere and is not an effect of aliasing. Figure \ref{fig:synthetic_zonally_shifted_lightcurve_no-K2_wave} shows a simulated light curve composed of two closely-spaced period components (4.7 and 5.25 hours) with a phase shift of 1.36 radians, with uniformed noise added. The figure shows that -- unlike as it was assumed in some past works --  the two waves do not lead to $k=2$ period component. We repeated this analysis for two-period components with the same value for around 5 hours and cycled through a number of phase shift values. We found that all the power at 2.5 hours was at least 2 orders of magnitude smaller than the power at the primary period of 5 hours. The $k=2$ periods suggest that Luhman 16 AB would thus have waves at half the period of $k=1$ waves.

\subsection{Planetary-scale Waves vs. Fourier Series}

{With k=1 and k=2 waves identified in our analysis, it may be relevant to contrast the planetary-scale waves model with the Fourier series. Superficially, it may appear that the success of the planetary-scale wave model is due to its mathematical similarity to the Fourier series. However, upon closer inspection, the two are substantially different, as explained below.}

{Despite the similarities, the planetary-scale wave model is mathematically different from a Fourier series expansion in two important ways. First, in the Fourier series, the function is expanded as a sum of waves with decreasing periods that form a series ($P$, ${P\over2}$,${P\over3}$,..., ${P\over n}$). In essence, as the number of series elements increases, the frequency space (periodogram) is increasingly covered. However, the planetary-scale wave model is different: Periods are not fixed absolute or relative values but are mostly unconstrained. They converge to two groups of periods (group of periods around k=1 and k=2), rather than to a Fourier series. This is significant because these waves do not cover a broad frequency space, but the opposite: They concentrate into 1 or 2 groups.}

{The second key difference we will highlight is that Fourier expansion is applicable to \textit{periodic} functions and the fundamental mode of the series (the longest period) is set to equal the entire length of the data. Therefore, the Fourier series is not used or should be expected to reproduce aperiodic functions beyond the extent of the fundamental mode (longest wavelength). This, in our case, for example, a Fourier expansion's periods would be $1,200$~h, $600$~h, 400~h, 240~h, 200~h, etc.} 

{In contrast, our planetary-scale wave models converge on a sum of sine waves with periods grouping around 2.5 and 5.2~h, i.e., about 200 times shorter periods than the fundamental modes of the Fourier series would be.}

{In short, the planetary-scale wave model fits the lightcurve with a sum of sine waves with periods as free parameters, which is similar but mathematically and functionally different from a series of sine waves with periods that decrease as $P \over n$.}

\subsection{Models of Gas Giant Windspeed Distribution}\label{subsec:}

Returning to Solar System gas giants, the comparison of Luhman 16 AB and Neptune's $k=1$ and $k=2$ highlighted an interesting difference: The period multiples of the strongest peak in the period distribution of Luhman 16 AB do not match, whereas these period multiples match relatively well in the Neptune data \citep[][]{simon_neptunes_2016}. In the next section, we will explore this difference.

Both Neptune's and Uranus' zonal jets are not strongly visible in wavelengths available to Voyager (see comprehensive review by \citealt{galperin_gas_2019}, \citealt{fletcher_how_2020}), although they show more structures and zonal banding in the NIR wavelength at epochs where cloud-top variability is greatest, i.e. Neptune \citep[][]{irwin_spectral_2023}; Uranus with Keck H-band \& Voyager 2 joint reanalysis \citep[][]{sromovsky_high_2015}. Most of Neptune's deeper tropospheric zonal circulation is not visible in Kepler photometry. The variability in \citet{simon_neptunes_2016} is dominated by the top-of-atmospheres cloud bands and not from waves coming from deeper zonal bands. 

In the case of Luhman 16 B, the difference in period multiples of $k=1$ and $k=2$ waves hint at two different ranges of period distribution which may contain spatial information. In order to explore why the period distribution may be narrower for the $k=2$ period group, we will use a simple model of Jupiter's period distribution as a function of latitude.  

\begin{figure*}
    \centering
    \includegraphics[width=0.99
\textwidth]{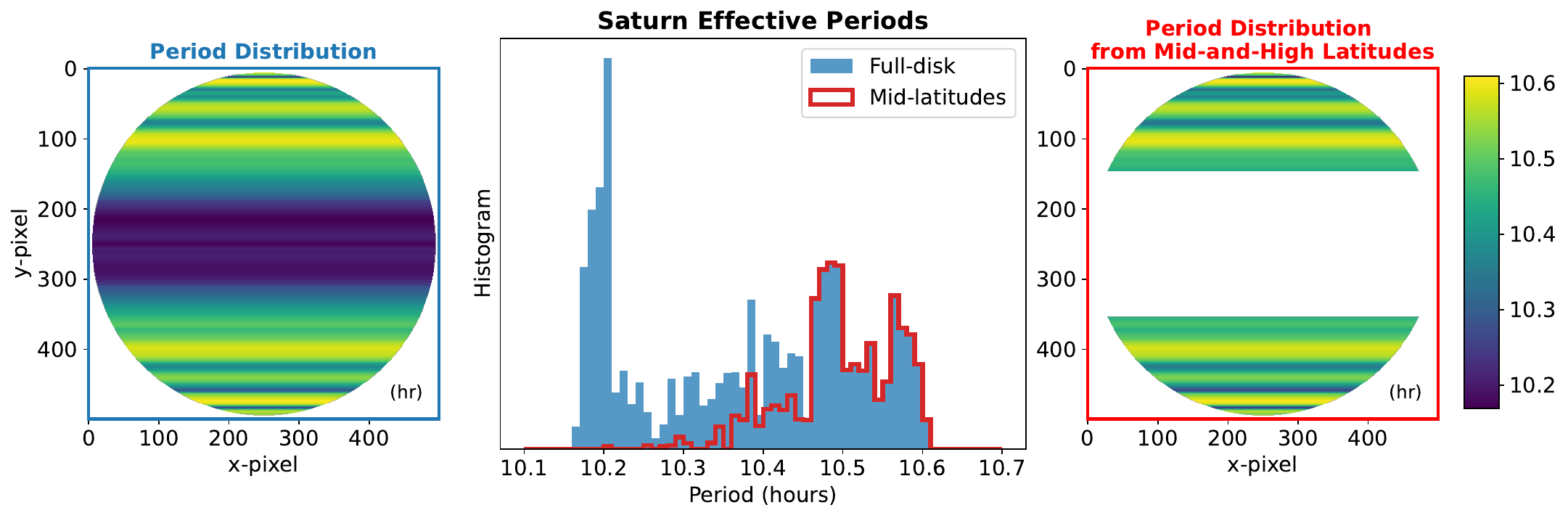}
    \caption{The synthetic period distribution of Saturn based on Cassini windspeed data (\citealt{garcia-melendo_saturns_2011}) as a function of latitude. \textit{\textbf{Left}}: Zero-inclination, synthetic image showing periods distribution of all latitudes. \textit{\textbf{Right}}: Zero-inclination synthetic image showing periods distribution only from latitudes above 35$^\circ$N and below 35$^\circ$S. \textit{\textbf{Center}}: Histograms of effective period for two cases.
    \label{fig:CassiniSaturn_model}}
\end{figure*}


{In our model of Jupiter, f}ollowing \citet[][]{apai_tess_2021}, in order to calculate the effective period of winds as a function of {latitude}, we {used} published windspeed measurements of Jupiter from the Cassini mission taken from \citet{porco_cassini_2003}. In a fly-by in 2003, Cassini took multiple images of Jupiter over one rotational period and tracked the clouds to measure windspeed. The revolution time for each latitude is the effective period at that latitude. 

From the wind speed measurements as a function of latitudes, we generated a synthetic model composed of the period-band structure. The effective period $P_\text{eff}$ is related to the latitudinal windspeed by:

$$P_\text{eff}(\theta)= \frac{L}{u_\text{rot}+u_\text{wind}}$$

where $L=2\pi r\cos(\theta)$ is the circumference at each latitude $\theta$, the rotational speed $u_\text{rot}=L/P$ where $P$ is Jupiter's self-rotation period, and $u_\text{wind}$ the measured windspeed at each latitude $\theta$. Thus we obtain an equirectangular projection of the period-band, which could be mapped back onto a sphere. This is achieved via the 3D mesh method in the visualization package \verb|mayavi|\footnote{\href{https://mayavi.readthedocs.io/en/latest/auto/example_spherical_harmonics.html}{\textbf{mayavi} Spherical Projection.}} \citep[][]{ramachandran_mayavi_2011} using a spherical mesh function. The final image used for analysis is an actual 3D sphere that is projected back onto a plane. This implementation provides more flexibility to explore the impact of different inclinations compared to the 2D model in \citet[][]{apai_tess_2021}.

Luhman 16 B is inclined at about $20^\circ$ (\citealt{apai_tess_2021}). We included a similar inclination in the 3D model of Jupiter shown in Figure \ref{fig:CassiniJupiter_model}. A different inclination changes the fraction of area visible between the polar and the equatorial regions, thus changing the period distribution in the visible disk.

We hypothesize that the k=2 period group may be emerging from spatially different distribution than the k=1 period group does. To test this, we masked out different regions of latitude and calculated the periodogram that emerges from the non-blocked regions. The results of this analysis and period histograms are shown in Figure \ref{fig:CassiniJupiter_model}. 

The highest and lowest wind speeds correspond to the shortest and longest effective periods, forming the tail ends of the full-disk (unmasked disk) period distribution. These shortest and longest period values are all located at the equatorial region within 35$^\circ$S to 35$^\circ$N. Hence, if we consider only the windspeed from the mid-latitudes to the polar region, excluding the equator, the period distribution is narrower than that of the full (unmasked) disk.


To broaden our windspeed comparison beyond Jupiter, we created a second period-band model of Saturn, assuming a zero-inclination, circular disk -- not taking into account the oblateness of Saturn.  We utilized the Cassini windspeed data as a function of latitude on Saturn (averaged from 2004 to 2009, \citealt{garcia-melendo_saturns_2011}) as shown in Figure \ref{fig:CassiniSaturn_model}. The Saturn measurements are averaged over a longer 5-year timescale. The resulting synthetic toy model at zero-inclination is shown in Figure \ref{fig:CassiniSaturn_model}.

The resulting period distribution of Saturn differs from Jupiter (Figure \ref{fig:CassiniJupiter_model}) in that Saturn's highest windspeeds (shortest periods) are all concentrated in the equatorial region, in contrast to the more varied windspeed values in the equatorial region of Jupiter. 

In the case of Saturn, almost the entire shorter-period half of the distribution is cut off in the masked disk, in contrast with Jupiter where only the tail ends are cut off. Nonetheless, this shows that our simple models are consistent in one way: Both their period distribution shrinks in the period range once the equatorial region is discarded, showing that the mid-latitude-to-polar region has a smaller range of period distribution.


 The narrowing of the period distributions at different latitudes in our Jupiter and Saturn model mirrors the difference in the period distributions of the $k=1$ and $k=2$ on Luhman 16 B. Thus, a natural explanation for the \luB{} pattern is that the $k=1$ and $k=2$ waves arise from different latitudes, such that the $k=2$ wavenumber arises from only the mid-latitude region.

\subsection{Short-period Light Curve Fits}
\label{subsec:discuss_lc_fit}

The fact that a simple model of multi-sine waves analogous to planetary-scale waves could fit the light curve over multiple timescales is a strong support of the rotation and 3D circulation driving zonal circulation on Luhman 16 AB. This result is in strong agreement with previous works \citep[][]{apai_tess_2021, apai_zones_2017-1} and shows the applicability of planetary-scale waves for certain brown dwarf atmospheres including Luhman 16 AB. 

We showed that the planetary-scale wave model is a simple model that could explain rotational modulation in the data over tens of time the rotational period, i.e. up to 100 hours for Luhman 16 AB.

\subsection{Long-Period Light Curve Evolution}
\label{subsec:discuss_longPeriodLC}

Via applying a frequency bandpass to isolate long-period varying components, we showed that the Luhman 16 AB light curve contains components that vary on timescales much longer than the rotational period, with periods up to 100 hours. In this section, we will discuss some potential mechanisms for long-period variability in brown dwarf atmospheres. 

A review of theoretical models and observations from \citet{showman_atmospheric_2020} suggests that long-term atmospheric variation is a trademark feature of gas giants and brown dwarfs. Quasi-periodic oscillation in the equatorial jets on a multi-year timescale exists on both Jupiter and Saturn, where eastward and westward jets migrate in the latitudinal direction and change their positions over time. This quasi-periodic oscillation has a period of about 4 years for Jupiter and 15 years for Saturn. The 3D GCMs for gas giants' atmospheres (\citealt{showman_atmospheric_2019}) also captured these oscillations with periods ranging from a few years up to 12 years.

Motivated by the unexpected long-term observed variability of Jupiter by \citet{orton_unexpected_2023}, \citet{hori_jupiters_2023}  demonstrated theoretically that the torsional oscillations arising from the rapid rotation of ionized, conducting cylindrical layers in the deep interior of Jupiter can explain the timescale of cloud-level variability observed with Juno. This torsional force between the interior layers bends the magnetic fields of Jupiter and creates waves that propagate to the cloud top. This is a potential model to explain the long-period, multi-year variability seen in Jupiter's atmosphere.

There is a clear timescale difference between the long-period oscillations in our Luhman 16 AB data, which go up to hundreds of hours, versus the multi-year timescale discussed in the literature. It could be speculated that the difference in mass, magnetic field strength, and the convective interior might play a role in timescale differences of long-term atmospheric variation between brown dwarfs and gas giants. Still, the nature of long-period variability remains inconclusive because there has not been a time-resolved, year-long monitoring dataset of brown dwarf atmosphere yet. Indication of long-term changes on brown dwarfs exists via comparing photometry taken several years apart (Luhman 16 B, \citealt{bedin_hubble_2017}), but it is unclear how these multi-year oscillations would arise on brown dwarf atmospheres and to what degree they are connected to shorter-period variation. In the near future, time-resolved extended multi-band monitoring will be the key to understanding the physics behind long-period atmospheric variation on brown dwarf atmospheres.

\subsection{Insights from GCMs for the explanation of short and long period light curve}\label{subsec:discuss_GCM}

\begin{figure*}[ht!]
\centering
\includegraphics[width=\textwidth]{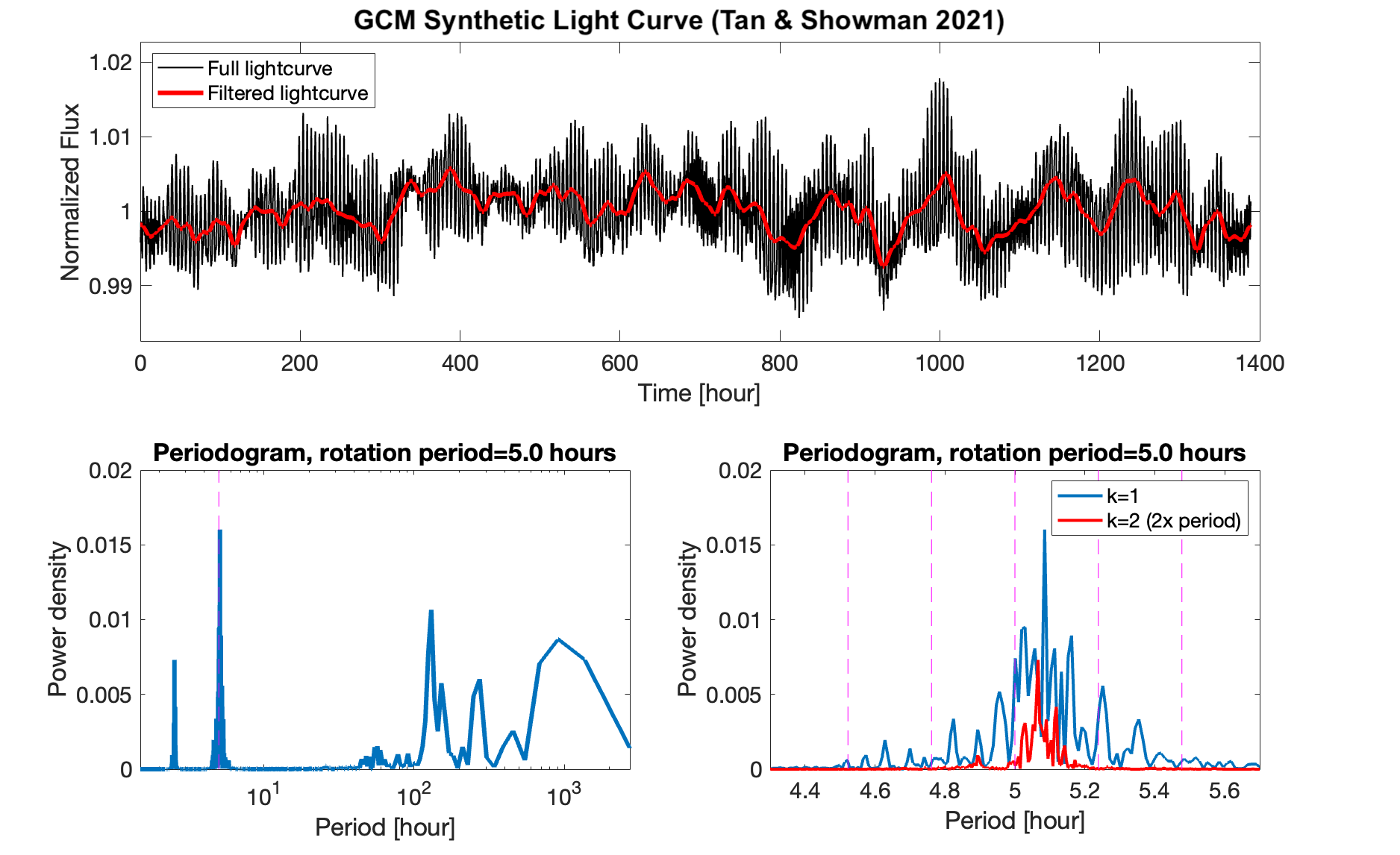}
\caption{Modelled light curve and periodogram power spectrum from a GCM presented in \citet{tan_atmospheric_2021}. The model behind this light curve assumes a rotation period of 5 hours and a moderate basal drag with a drag timescale  $\tau_{\rm drag} = 10^6$ s, such that meridionally broad westward and eastward jets with a peak speed of $\sim 500$ ${\rm ms^{-2}}$ formed in the model. \textit{\textbf{ Upper panel}}: modeled light curve spanning over 1300 hours with a sampling interval of 10 minutes which is the same as that of the TESS dataset in this work. The black line is the original light curve and the red line is the filtered light curve in which signals with periods less than 15 hours are filtered out.  \textit{\textbf{Lower left}}: Periodogram power spectrum of the modeled light curve, and the dashed line highlights the underlying rotation period. \textit{\textbf{ Lower right}}: Periodogram power spectrum zooming into the vicinity of the rotation period (blue line), and the red line is the power spectrum with twice of the period, highlighting the $k=2$ component. Dashed lines from left to right represent signals  Doppler-shifted by a jet with eastward velocities of 2, 1, 0, -1, and -2 ${\rm km~s^{-1}}$.  Both the $k=1$ and $k=2$ components show redshifts compared to the rotation period, and this is because of the Doppler shift by the broad, westward equatorial jet formed in the model. }
\label{fig:TS2021model}
\end{figure*}

GCMs of brown dwarf atmospheres in \citet{tan_atmospheric_2021} demonstrated that cloud radiative feedback generates east-west traveling large-scale waves and stochastic vortices. With a weak basal drag which crudely represents interactions between the active weather layer and the deep quiescent interior, strong zonal jets can form and the jet speed depends on the drag strength. The inherent inhomogeneous cloud coverage naturally leads to rotational light curves. In addition, traveling waves and the statistical evolution of storms produce irregular features in the light curve which can broaden the power spectrum of the light curve around the rotation period; the presence of zonal $k=2$ waves imprints strong $k=2$ signatures in the light-curve power spectrum.

The modeled light curve and its power spectrum from a GCM in \citet{tan_atmospheric_2021} are presented in Figure \ref{fig:TS2021model}. This particular model assumes a rotation period of 5 hours, similar to that of Luhman 16B, and a moderate basal drag with a drag timescale of  $\tau_{\rm drag} = 10^6$ s such that the model developed a broad westward equatorial jet and high-latitude eastward jets with a peak speed of $\sim 500$ ${\rm ms^{-2}}$ (see Figure 9 in \citealp{tan_atmospheric_2021}). The modeled light curve shows both high-frequency variations that are caused by the rotational modulation of the zonal $k=1$ and $k=2$ atmospheric features and low-frequency variations that vary over timescales on the order of 100 hours. Powers near 5 hours are significantly broadened and discretized into multiple peaks that can extend to $\pm 0.4$ hours away from the major peak, quantitatively in good agreement with the observed power spectrum shown in Figure \ref{fig:periodograms_k1-Andk2}. The $k=2$ power in the modeled light curve shows a narrower broadening than the $k=1$ components, similar to the observed one; however, this model did not produce double $k=2$ groups which is robustly seen in the observed one. The longer-timescale variability is from the statistical evolution of the ensemble storms, despite that individual storms could have a much shorter evolution timescale. This statistical fluctuation has been shown in box models that exclude equatorial waves and rotational modulations in \citet{tan_atmospheric_2021}. The modeled long-term variability is remarkably similar to the observed one for Luhman 16B despite that the model was not tuned for Luhman 16B.

Note that zonal jets in the model are not the main reason for the broadened power spectrum. The equatorial westward jet provides a redshift for the bulk power spectrum including the $k=1$ and $k=2$ components. High-latitude eastward jets  contribute little to the power spectrum because low-latitude variability dominates the light curve.  A similar model with a strong drag generates a similarly broadened power spectrum but no redshift; another similar model with an even weaker drag (therefore with a stronger westward equatorial jet) produces a similar power spectrum that is merely red-shifted more.

Comparisons between the observed and modeled light-curve power spectra indicate a somewhat alternative view for some features of the observed power spectrum of Luhman 16B: Luhman 16B may potentially develop a global storm system in the atmosphere that host self-organized equatorial waves and vortices driven primarily by the cloud radiative feedback; $k=1$ light-curve components are broadened by the traveling waves and stochastically evolving storms, while longer statistically fluctuation of the storm system drive long-term light-curve variability.

Critically, the presence of two $k=2$ groups in the observed power spectrum (but lacks in the models) suggests the development of alternating jets on Luhman 16B that are capable of affecting the observed light curve. In the model framework of \citet{tan_atmospheric_2021}, this indicates that the alternating jets should form at lower latitudes where most variability is from, necessitating further modeling efforts to understand jet formation on Luhman 16B.


\subsection{Potential Physical Interpretation}

{At this point, despite its success in explaining the available rotational modulations over many rotational periods, the planetary-scale wave model is a mathematical construct providing a simple phenomenological model. It is not a self-consistent physical model, although it is physically motivated and it aligns with the general predictions of detailed dynamical models (see Section~\ref{subsec:discuss_GCM}). In the following, we will provide a potential physical interpretation without claiming this to be a complete (or even unique) interpretation.}

{Our interpretation emerges from a combination of four established facts: (1) Atmospheric waves are present in all atmospheres and their potential wavelengths range from molecular scales to planetary scales. (2) Rotationally-dominated, internally heated gaseous atmospheres will develop zonal circulation, differential rotation, and latitudinally varying wind speeds. (3) Condensation products (dust particles) form in brown dwarf atmospheres at locations where the pressure-temperature conditions lead to super-saturation. (4) The L/T spectral type transition in brown dwarfs is broadly interpreted to occur as the top of the silicate cloud deck just aligns/approaches the optical depth probed by observers.} 

{By combining the four facts above and following them to their conclusion, we provide the following potential physical interpretation for the planetary-scale waves in Luhman 16B: In this object, we observe the top of the cloud deck close to the observable optical depth. Small-scale turbulence and mixing drive waves of increasing wavelengths, all the way to the longest-possible wavelength standing waves with wavelength equaling the circumference of the object. As the circumference changes with latitude and the rotating gaseous also inhibits differential rotation, the longest wavelength and the rotational period corresponding to the longest waves will vary somewhat with latitude. Waves with vertical components will periodically move gas parcels up and down in the atmosphere. Such motion close to the saturation curve of the key condensates (silicates and iron) will result in periodic condensation/evaporation of these, similar to gravity-wave-induced spatially periodic water vapor clouds that are common in Earth's atmosphere. }

{Put together, the planetary-scale waves will introduce a spatially periodic modulation in cloud thickness, that will be observed to show a temporal periodicity with a range of periods similar to the overall average rotation period of the atmosphere, but showing variations due to differential rotation and wind speeds. 
Our data also shows very strong evidence for standing waves that correspond to half the circumference of the atmosphere, also showing multiple peaks such as the k=1 waves.} 

{The above qualitative interpretation appears to match the body of evidence we are aware of, and aligns well with the general findings of state-of-the-art GCMs, without requiring an assumption of any individual process or phenomenon that is not already established. }

{Perhaps the only seemingly surprising aspect of this interpretation may be the assumption that the large-scale waves coincide in their vertical location with the cloud top (i.e., pressure-temperature conditions that are required for super-saturation of silicates). Generally and in most atmospheres, there may not be an obvious physical reason for these two to spatially coincide. However, this apparent contradiction is readily resolved when one considers the target selection biases: Most rotational variability studies focus on L/T transition brown dwarfs because that is the spectral type range where the largest-amplitude modulations are observed (e.g., \citealt{radigan_independent_2014} -- which observation then naturally aligns with the explanation above. (Luhman 16B is specifically targeted because it is one of the most variable brown dwarfs.) In other words, in atmospheres where the large-scale waves are well above or well beneath the condensate cloud deck's top, the waves will \textit{not} modulate the cloud properties; \textit{not} break the axisymmetry of the atmosphere, nor leading to large-amplitude rotational modulations.}

{While the above interpretation provides a compelling and qualitatively consistent picture, we caution that it is not yet a quantitative and self-consistent model. While individual model components exist and support the above picture, more work is needed to build a multi-component, self-consistent model that can provide a quantitative model for the planetary-scale waves' impact on cloud thickness in L/T brown dwarfs.}

\section{Conclusions}\label{sec:Conclusion}

Luhman 16 AB is the brightest brown dwarf binary observable from the Solar System and an excellent target for probing ultra-cool atmospheres. This observation helps to understand the broad parameter space of atmospheres to which ultra-cool atmospheres like those of Jovian planets and sub-Neptunian planets also belong. Below we summarized our findings of Luhman 16 AB photometric monitoring.

\begin{enumerate}
    \item \textit{Longest photometric monitoring of Luhman 16 AB}: With light curve data covering 1,200 hours (50 days) baseline with a 10-minute cadence, we found evidence for strong, persistent photometric variability with an amplitude of about $\pm5\%$. We found strong periodic modulations around the rotational period of Luhman 16 B centered at $\sim$5 hours and half the rotational period at $\sim$2.5 hours. We ruled out the likelihood of photometric contamination by spacecraft pointing drifts or background sources. {We also ruled out contribution from Luhman 16 A around 7.5 hours, due to very low variability amplitude in our dataset around this period.}
    
    \item \textit{Complex structure in short-period (P$<$10~h) periodogram}: Our generalized Lomb-Scargle periodogram analysis shows a range of amplitudes and periods, consistent with previous work \citep[][]{apai_tess_2021}. Short-period variability (P$<$6~h)  originates from rotational modulations in the atmosphere of Luhman 16 B. We showed that:
    \begin{enumerate}
     \item Strong power exists at periods near the rotational period of Luhman 16 B: Both in the primary $k=1$ periods in the range 4.25--5.75 hours with peaks at 4.7 and 5.25 hours, and the $k=2$ periods in the range 2.3--2.65 hours with peaks at 2.43 and 2.575 hours.
     \item A multi-sine model analogous to planetary-scale waves could fit the periodogram of Luhman 16~B centered at 5 hours;
     \item There is a multi-peak patterns coinciding with the $k=2$ (half rotational period) range. The pattern is similar to the $k=1$ periods. Via analysis of Jupiter and Saturn's effective period distribution and Neptune's power spectra, we show that the observed distribution is observed if the k=2 waves are mostly present in mid-to-high latitudes, while k=1 waves are also present in the equatorial regions,
    \end{enumerate}
    
    \item \textit{The planetary-scale wave model explains the light curves}: The sum of 3--4 sine waves provides a very good match to the light curves (1$\sigma$ residuals $\sim$1\%). We found that the period distributions in our fits agreed well with the dominant components in the periodogram of Luhman 16 B. The periods converge around 2.5 hours and 5 hours for all light curve segments -- similar to the $k=1$ and $k=2$ periods. Our fits also showed that the periods and amplitude of sine waves experience time evolution as we move from one segment to another.

    \item \textit{Sustained Long-period variability}: We found long-period variability (different from rotational modulations) with an amplitude of about $5\%$ (much longer than the $\sim$5.2~hour rotational period of Luhman 16~B). 
    The long-period components likely originate from distinct regions in the atmosphere, independent of short-period zonal circulation.

    \item \textit{Qualitative consistency with GCM results}: We found that certain properties in the period distribution of our observations are consistent with specific simulations of weak-drag, rapidly rotating atmospheres from \citet{tan_jet_2022} and \citet{tan_atmospheric_2021}. Comparing the data and simulations, both the $k=1$ and $k=2$ period components are found in the periodogram, as well as the narrowed period distribution between $k=1$ and $k=2$ periods. Long-period variation in the light curve of up to hundreds of hours is also found.
    
\end{enumerate}
Our study demonstrates the power of very long-baseline photometric monitoring of brown dwarf and exoplanet atmospheres and opens a window on brown dwarf brightness evolution over tens and hundreds of rotations. Further high-cadence data will allow the k=2 waves described here to be studied in greater detail.

\newpage    
\section*{Acknowledgements}
Support for program number HST-GO-16296 was provided by NASA through a grant from the Space Telescope Science Institute, which is operated by the Association of Universities for Research in Astronomy, Incorporated, under NASA contract NAS5-26555. D.N. and L.R.B. acknowledge support by MIUR under PRIN program \#2017Z2HSMF and by PRIN-INAF 2019 under program \#10-Bedin. This paper includes data collected by the \textit{TESS} mission, which are publicly available from the Mikulski Archive for Space Telescopes (MAST). Funding for the TESS mission is provided by NASA’s Science Mission directorate.
All the {\it TESS} data used in this paper can be found in MAST: \dataset[10.17909/ndj7-4v42]{http://dx.doi.org/10.17909/ndj7-4v42}.

\section*{Author Contribution}
N.F. led the data analysis and modeling, prepared the figures, and implemented new algorithms and software solutions.  D.A. conceived the overall project and provided programs from previous work that formed the basis of the analysis. N.F. and D.A. led the interpretation of the results. D.N. provided the processed PATHOS-extracted TESS light curves and contributed helpful insights on testing the background contamination levels. X.T. provided the synthetic light curves and periodograms from GCMs and contributed theory insights. T.K. and L.R.B. provided insightful input on the interpretation of rotational modulations. N.F. outlined and wrote the draft of the manuscript; D.A. wrote the introduction and contributed revisions throughout the manuscript. X.T. wrote the discussion of the GCM results. All authors participated in proofreading and provided comments and improvements to the manuscript.

\appendix
\section{Assessment of Contamination from Background Source}
\label{appendix:bg_source}

\begin{figure}
    \centering
    \includegraphics[width=0.45\textwidth]{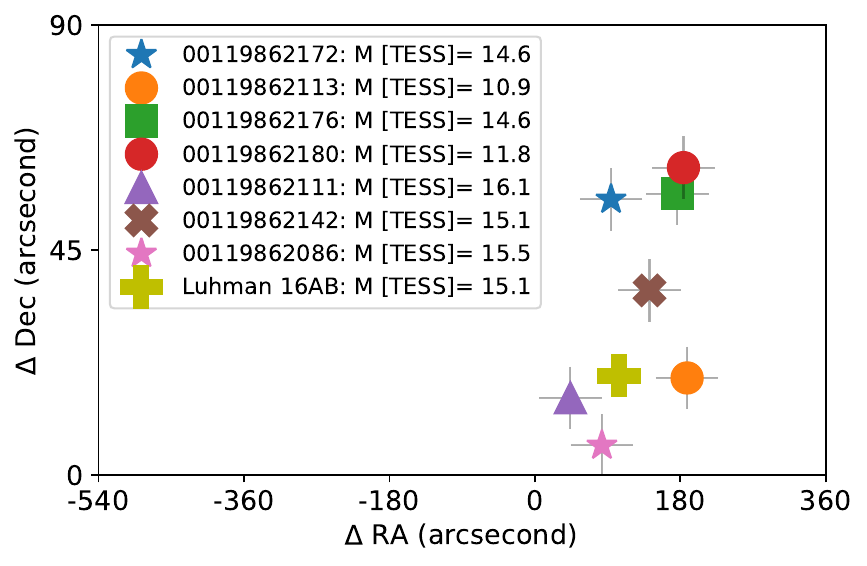}
    \caption{TESS sources within 2 arc-minutes of Luhman 16 AB with comparable or brighter magnitude, assessed for potential contaminants. TESS Identification Number and TESS-magnitude of each source are included in the legend.}
    \label{fig:appendix_bgSource_sky}
\end{figure}

Precursor TESS study of  Luhman 16 AB used high-resolution, multi-epoch Hubble Space Telescope data to effectively rule out photometric contamination from background sources \citep{apai_tess_2021}. In the following section, we again show that it is unlikely Luhman 16 AB contains photometric contamination from background sources in our data -- specifically when using the PSF-extracted light curve where the photometric-extraction radii are less than 1 TESS-pixel (about 21 arcseconds).

\begin{figure*}
    \includegraphics[width=0.49\textwidth]{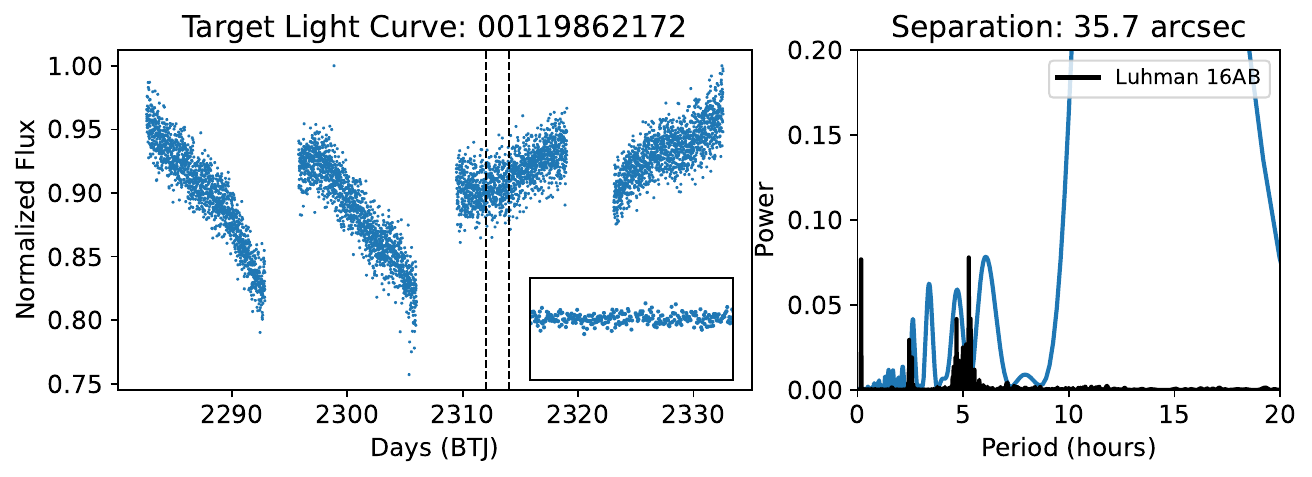}
    \includegraphics[width=0.49\textwidth]{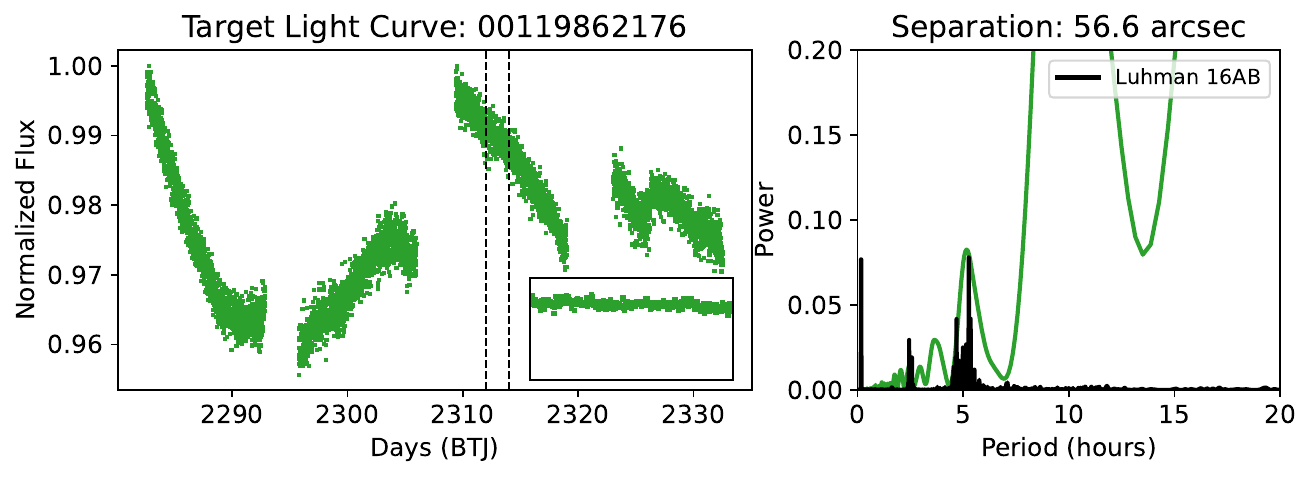} 
    \includegraphics[width=0.49\textwidth]{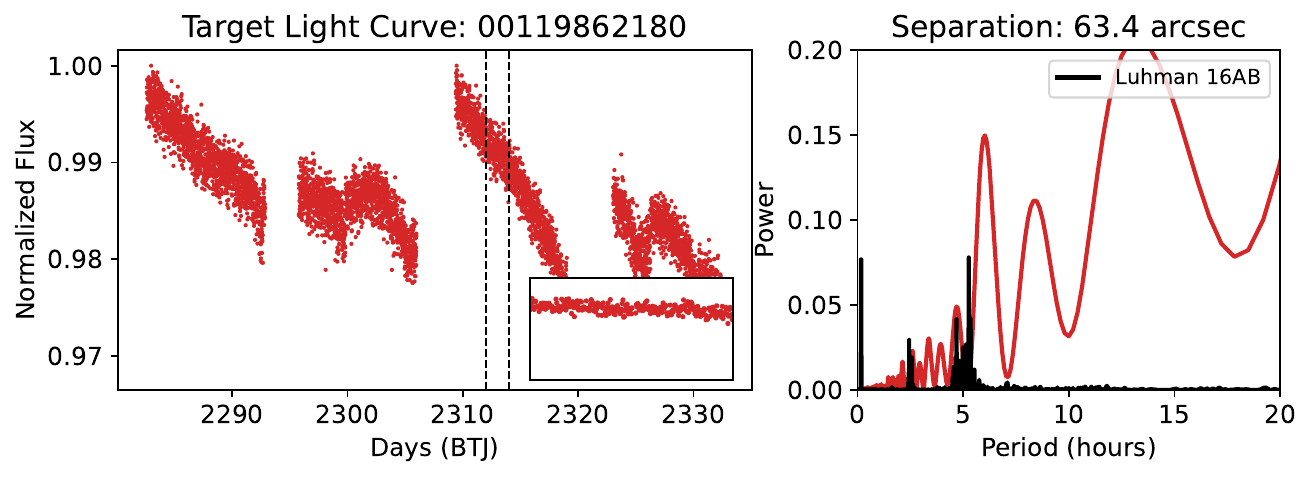}
    \includegraphics[width=0.49\textwidth]{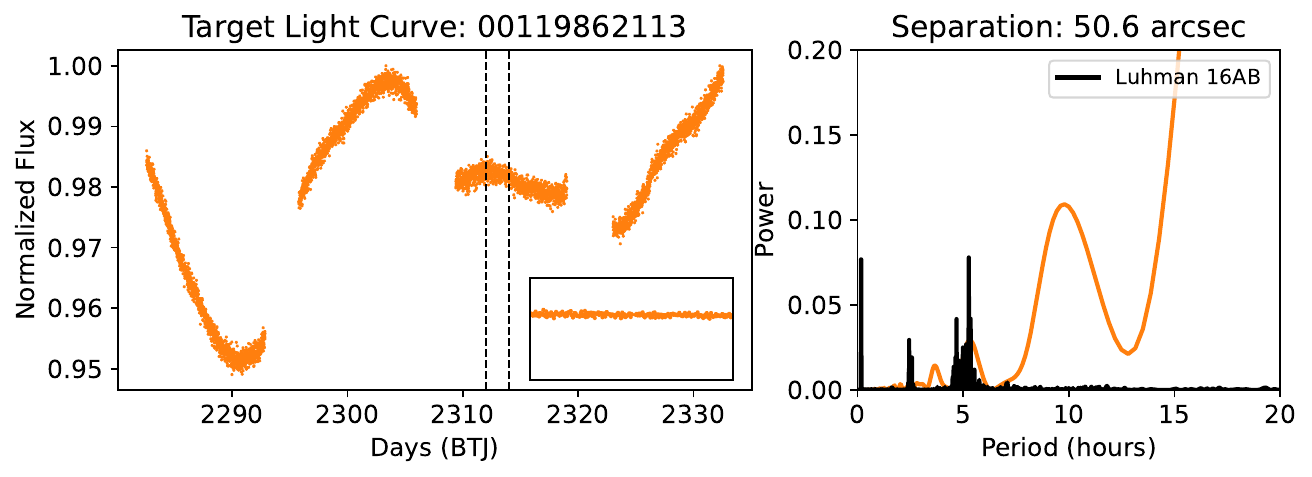}
    \includegraphics[width=0.49\textwidth]{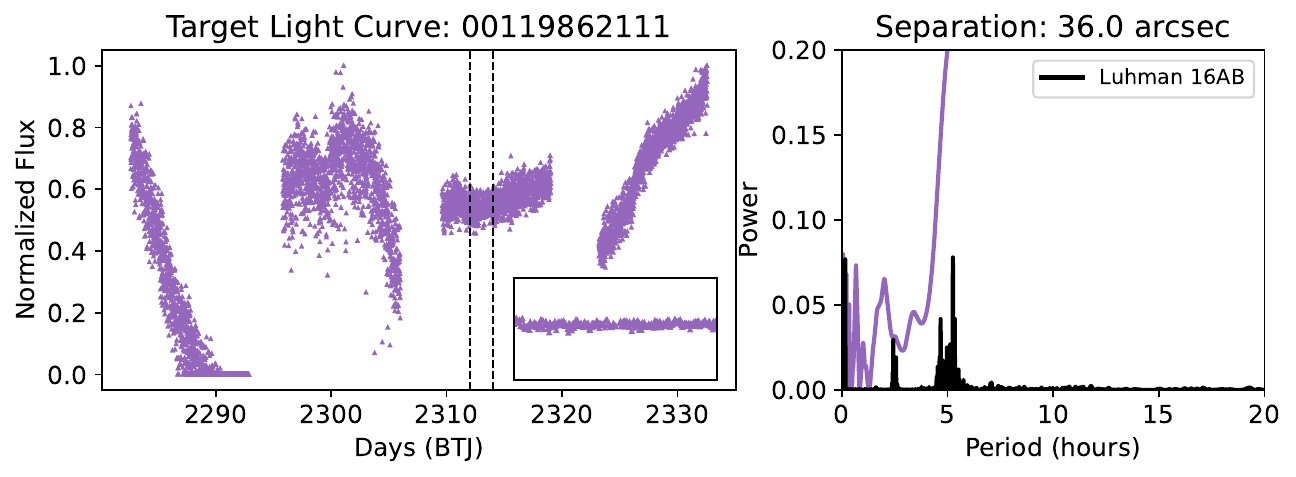}
    \includegraphics[width=0.49\textwidth]{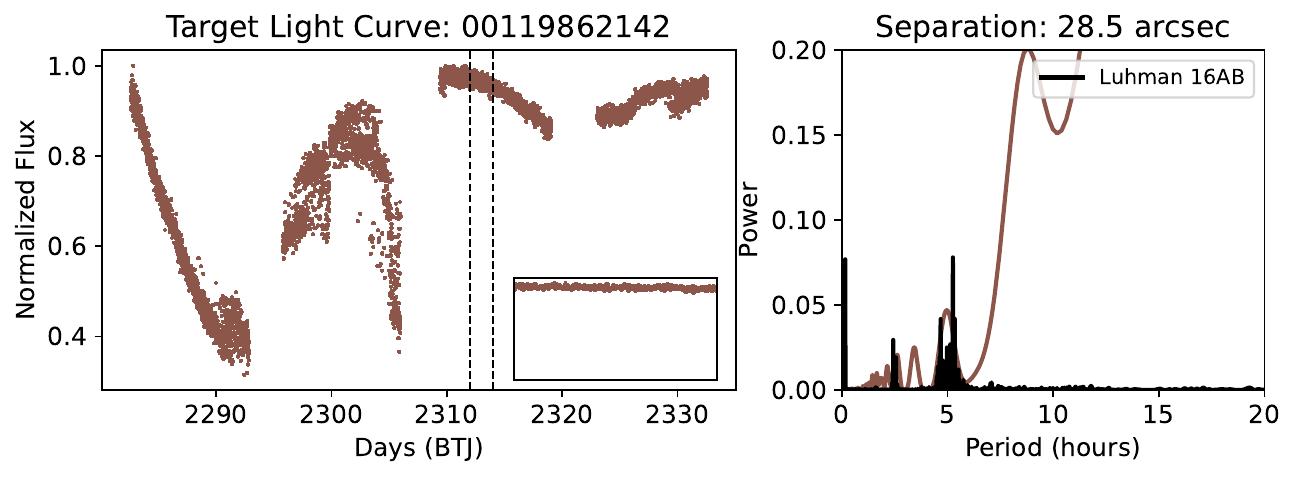}
    \includegraphics[width=0.49\textwidth]{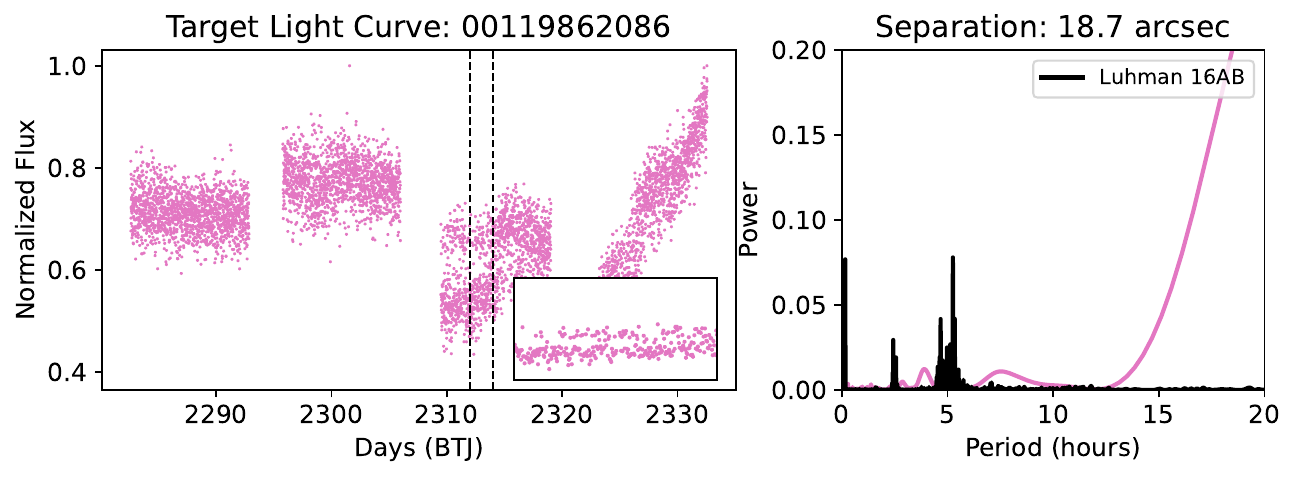}
\caption{Light curves and periodograms of background sources shown in Figure \ref{fig:appendix_bgSource_sky}. {One background source with separation under 21 arc-second (the TESS pixel size) from Luhman 16AB has a featureless light curve with no periodogram power around 5 hours -- where Luhman 16 AB periodogram power is strongest. The other sources with non-flat period power around 5 hours also have featureless light-curve and is located faraway from Luhman 16 AB (beyond the 21 arcsec limit)}. The inset plot shows a 48-hour window for each light curve.}
\label{fig:appendix_bgLightcurve}
\end{figure*}

In Figure \ref{fig:appendix_bgSource_sky}, we plot background sources of comparable or brighter TESS-magnitude (magnitude for TESS bandpass from 600--1000nm) than Luhman 16 AB within 2 arc-minutes. In Figure \ref{fig:appendix_bgLightcurve}, we plot the light curve and the resulting Lomb-Scargle periodograms for each of these objects. It can be seen that most objects shown do not contain strong period power under 5 hours where Luhman 16 AB variability is strongest. {It is noteworthy that the periodogram power shown via the Lomb-Scargles periodogram does not represent absolute power but rather normalized power, which is important when comparing periodograms from two different objects. Thus, an inspection of the light curve is always necessary to assess whether actual variability exists in the data, not just solely from whether certain period power exists in the periodogram.}

The closest source from Luhman 16 AB, TESS ID $00119862086$ with separation 18.7 arcsecs, has a featureless light curve with no periodogram power under 10 hours. The largest periodogram power under 20 hours is from source 00119862172 with a separation of 35.7 arcsecs, but this source also has a featureless light curve and is outside of Luhman 16 AB PSF-radii. Overall, we conclude that background source contamination of Luhman 16 AB is not a concern within our data set and our chosen photometric extraction parameters, consistent with rigorous examination of background sources from previous work of \citet{apai_tess_2021}.

\section{Assessment of over-fitting: three-sines versus four-sines model}
\label{appendix:3sine_bic}

\begin{figure*}
    \includegraphics[width=0.95\textwidth]{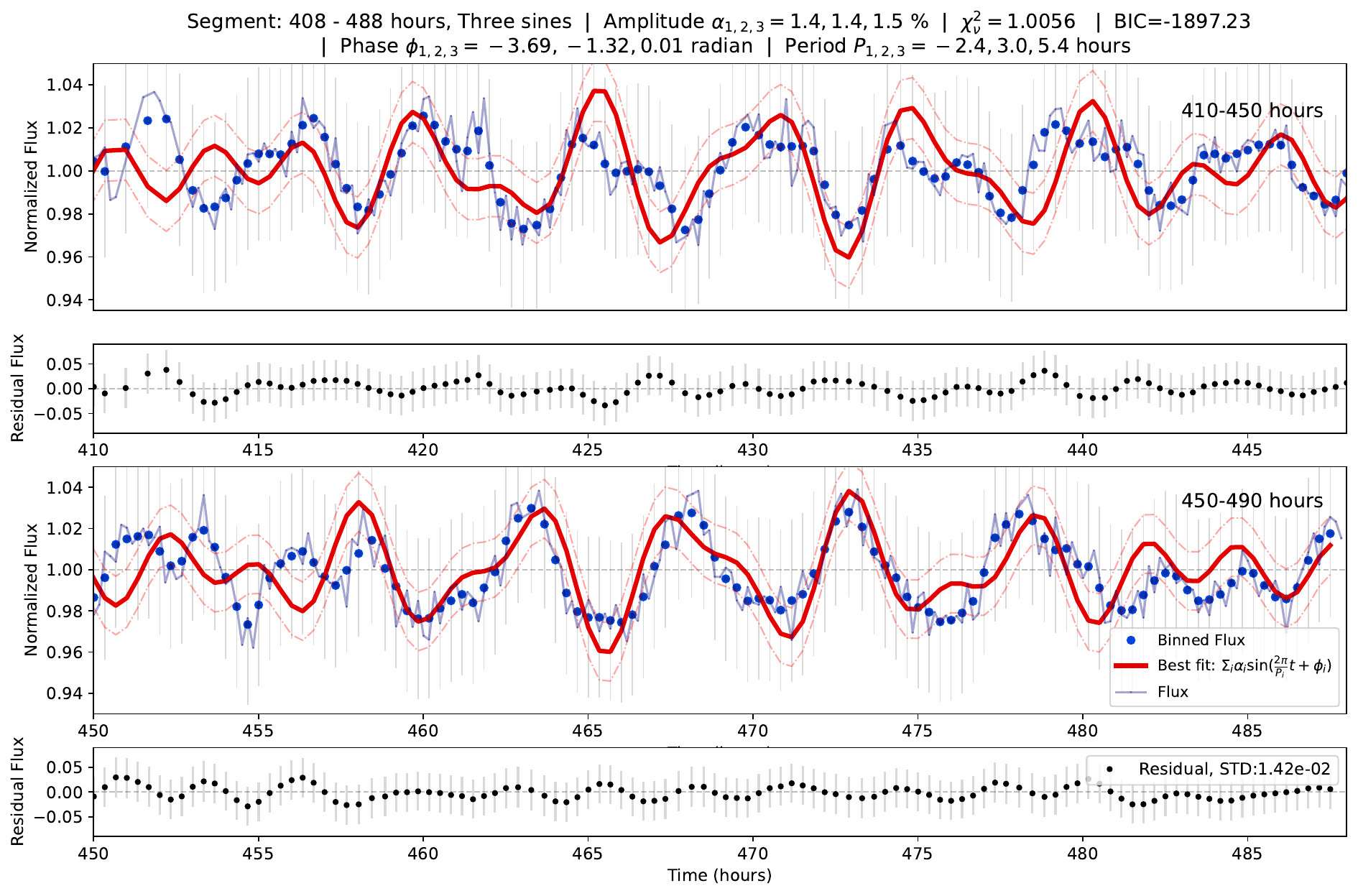}
    \caption{{The light curve fit of the 410-490 hours segment using the three-sine model fit (red curve) corresponding to the data (blue points). Thin vertical lines show the photometric error. Amplitude, period, phase information, the reduced chi-squared ($\chi_\nu^2=1.0056$), and the Bayesian Information Criterion score (BIC=$-1897.23$) are displayed in the title. Visually, the goodness of fit for the three-sine model is worse than the four-sine model. The three-sine model both has a higher reduced chi-squared and a higher BIC score compared to the four-sine model, indicating that the four-sine model fits the data better. For the result of the four-sine fit for the same data segment, see Figure \ref{fig:MCMC_4sines_sector3637_408_488-hr_fitpanelB}.}}
    \label{fig:hyperOPT_3sines_sector3637_408_488-hr}
\end{figure*}

{As mentioned in Section \ref{subsec:LCFitresult}, we used a four-sine model because a three-sine model did not provide a good fit. The result of the four-sine fit for the 410-490 hours data segment is shown in Figure \ref{fig:MCMC_4sines_sector3637_408_488-hr_fitpanelB}. The result of the three-sine fit for the same data segment is shown in Figure \ref{fig:hyperOPT_3sines_sector3637_408_488-hr}. The three-sine model could not capture all the components in the light curve in the same way that the four-sine model could. A quantitative comparison verifies this assessment: The four-sine model produced a lower reduced chi-squared compared to the three-sine model: $\chi_\nu^2=0.7412$ and $\chi_\nu^2=1.0056$ for the four-sine and three-sine models, respectively.}

{However, adding another sine-wave means an extra three parameters to the fit, thus increasing the risk of over-fitting the data. Therefore,  to ensure that we identify a model that describes the light curve behavior well while remaining simple. } 

{To quantify the goodness of fit and the risk of over-fitting (when using the four-sine versus three-sine model) for this 80-hour data segment, we used the Bayesian Information Criterion (BIC). (A formal description of the BIC score could be found in \citealt{bauldry_structural_2015}). BIC is commonly applied to model selection and takes into account the data, the fit, the total number of observations as well as the total number of fitted parameters.  BIC penalizes over-fitting the data. A lower BIC score indicates a better model, with a difference in BIC of 10 commonly interpreted as strong evidence that the model with lower BIC is significantly better \citep{bauldry_structural_2015}. }

We calculated BIC with the package \verb|RegscorePy|\footnote{RegscorePy: \href{https://github.com/UBC-MDS/RegscorePy}{github-RegscorePy} (MIT License)}. {Both the BIC and $\chi_\nu^2$ can be found in the figure title of Figure \ref{fig:MCMC_4sines_sector3637_408_488-hr_fitpanelB} and \ref{fig:hyperOPT_3sines_sector3637_408_488-hr}. The BIC score for the three-sine model in Figure \ref{fig:hyperOPT_3sines_sector3637_408_488-hr} is BIC$=-1897.23$. The BIC score for the four-sine model in Figure \ref{fig:MCMC_4sines_sector3637_408_488-hr_fitpanelB} is BIC$=-1954.26$. Thus, the BIC score is lower for the four-sine model and suggests the four-sine model fits the 80-hour data segment better while remaining simple.}

\bibliography{main.bib}{}
\bibliographystyle{aasjournal}
\end{document}